\documentclass[twoside,twocolumn,9pt]{article}
\usepackage{extsizes}
\usepackage[super,sort&compress,comma]{natbib} 
\usepackage[version=3]{mhchem}
\usepackage[left=1.5cm, right=1.5cm, top=1.785cm, bottom=2.0cm]{geometry}
\usepackage{balance}
\usepackage{mathptmx}
\usepackage{sectsty}
\usepackage{graphicx} 
\usepackage{lastpage}
\usepackage[format=plain,justification=justified,singlelinecheck=false,font={stretch=1.125,small,sf},labelfont=bf,labelsep=space]{caption}
\usepackage{float}
\usepackage{fancyhdr}
\usepackage{fnpos}
\usepackage[english]{babel}
\addto{\captionsenglish}{%
  
}
\usepackage{array}
\usepackage{droidsans}
\usepackage{charter}
\usepackage[T1]{fontenc}
\usepackage[usenames,dvipsnames]{xcolor}
\usepackage{setspace}
\usepackage[compact]{titlesec}
\usepackage{hyperref}
\usepackage{siunitx}
\usepackage{booktabs}
\usepackage{amsmath}
\usepackage{subcaption}
\usepackage[flushleft]{threeparttable}
\usepackage{tabularx}
\usepackage[title]{appendix}

\usepackage{epstopdf}

\definecolor{cream}{RGB}{222,217,201}

\begin{document}

\pagestyle{fancy}
\thispagestyle{plain}
\fancypagestyle{plain}{
\renewcommand{\headrulewidth}{0pt}
}

\makeFNbottom
\makeatletter
\renewcommand\LARGE{\@setfontsize\LARGE{15pt}{17}}
\renewcommand\Large{\@setfontsize\Large{12pt}{14}}
\renewcommand\large{\@setfontsize\large{10pt}{12}}
\renewcommand\footnotesize{\@setfontsize\footnotesize{7pt}{10}}
\makeatother

\renewcommand{\thefootnote}{\fnsymbol{footnote}}
\renewcommand\footnoterule{\vspace*{1pt}%
\color{cream}\hrule width 3.5in height 0.4pt \color{black}\vspace*{5pt}} 
\setcounter{secnumdepth}{5}

\makeatletter 
\renewcommand\@biblabel[1]{#1}            
\renewcommand\@makefntext[1]%
{\noindent\makebox[0pt][r]{\@thefnmark\,}#1}
\makeatother 
\renewcommand{\figurename}{\small{Fig.}~}
\sectionfont{\sffamily\Large}
\subsectionfont{\normalsize}
\subsubsectionfont{\bf}
\setstretch{1.125} 
\setlength{\skip\footins}{0.8cm}
\setlength{\footnotesep}{0.25cm}
\setlength{\jot}{10pt}
\titlespacing*{\section}{0pt}{4pt}{4pt}
\titlespacing*{\subsection}{0pt}{15pt}{1pt}

\fancyfoot{}
\fancyfoot[LO,RE]{\vspace{-7.1pt}\includegraphics[height=9pt]{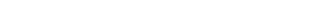}}
\fancyfoot[CO]{\vspace{-7.1pt}\hspace{13.2cm}\includegraphics{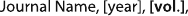}}
\fancyfoot[CE]{\vspace{-7.2pt}\hspace{-14.2cm}\includegraphics{head_foot/RF}}
\fancyfoot[RO]{\footnotesize{\sffamily{1--\pageref{LastPage} ~\textbar  \hspace{2pt}\thepage}}}
\fancyfoot[LE]{\footnotesize{\sffamily{\thepage~\textbar\hspace{3.45cm} 1--\pageref{LastPage}}}}
\fancyhead{}
\renewcommand{\headrulewidth}{0pt} 
\renewcommand{\footrulewidth}{0pt}
\setlength{\arrayrulewidth}{1pt}
\setlength{\columnsep}{6.5mm}
\setlength\bibsep{1pt}

\makeatletter 
\newlength{\figrulesep} 
\setlength{\figrulesep}{0.5\textfloatsep} 

\newcommand{\topfigrule}{\vspace*{-1pt}%
\noindent{\color{cream}\rule[-\figrulesep]{\columnwidth}{1.5pt}} }

\newcommand{\botfigrule}{\vspace*{-2pt}%
\noindent{\color{cream}\rule[\figrulesep]{\columnwidth}{1.5pt}} }

\newcommand{\dblfigrule}{\vspace*{-1pt}%
\noindent{\color{cream}\rule[-\figrulesep]{\textwidth}{1.5pt}} }

\makeatother

\twocolumn[
  \begin{@twocolumnfalse}
{\includegraphics[height=30pt]{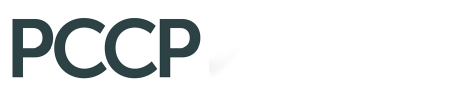}\hfill\raisebox{0pt}[0pt][0pt]{\includegraphics[height=55pt]{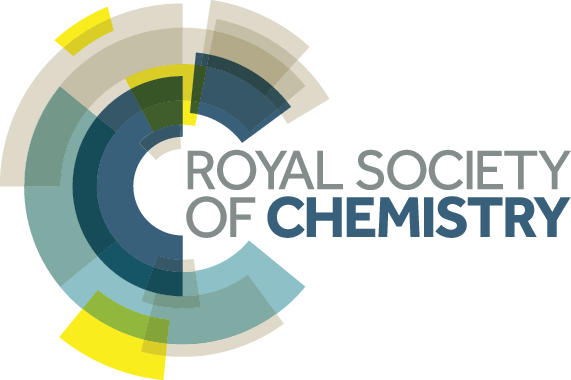}}\\[1ex]
\includegraphics[width=18.5cm]{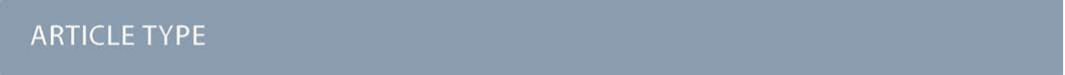}}\par
\vspace{1em}
\sffamily
\begin{tabular}{m{4.5cm} p{13.5cm} }

\includegraphics{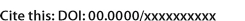} & \noindent\LARGE{\textbf{Sensitivity Analysis of Aromatic Chemistry to Gas-Phase Kinetics in a Dark Molecular Cloud Model$^\dag$}} \\
\vspace{0.3cm} & \vspace{0.3cm} \\

 & \noindent\large{Alex N. Byrne,$^{\ast}$\textit{$^{a}$} Ci Xue,\textit{$^{a}$} Troy Van Voorhis\textit{$^{a}$}, and Brett A. McGuire$^{\ast}$\textit{$^{ab}$}} \\

\includegraphics{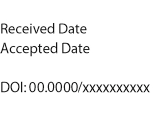} & \noindent\normalsize{The increasingly large number of complex organic molecules detected in the interstellar medium necessitates robust kinetic models that can be relied upon for investigating the involved chemical processes. Such models require rate constants for each of the thousands of reactions; the values of these are often estimated or extrapolated, leading to large uncertainties that are rarely quantified. We have performed a global Monte Carlo and a more local one-at-a-time sensitivity analysis on the gas-phase rate coefficients in a 3-phase dark cloud model. Time-dependent sensitivities have been calculated using four metrics to determine key reactions for the overall network as well as for the cyanonaphthalene molecule in particular, an important interstellar species that is severely under-produced by current models. All four metrics find that reactions involving small, reactive species that initiate hydrocarbon growth have large effects on the overall network. Cyanonaphthalene is most sensitive to a number of these reactions as well as ring-formation of the phenyl cation (\ce{C6H5+}) and aromatic growth from benzene to naphthalene. Future efforts should prioritize constraining rate coefficients of key reactions and expanding the network surrounding these processes. These results highlight the strength of sensitivity analysis techniques to identify critical processes in complex chemical networks, such as those often used in astrochemical modeling.} \\

\end{tabular}

 \end{@twocolumnfalse} \vspace{0.6cm}

  ]

\renewcommand*\rmdefault{bch}\normalfont\upshape
\rmfamily
\section*{}
\vspace{-1cm}


\footnotetext{\textit{$^{a}$~Department of Chemistry, Massachusetts Institute of Technology, Cambridge, MA 02139, USA; E-mail: lxbyrne@mit.edu, brettmc@mit.edu}}
\footnotetext{\textit{$^{b}$~National Radio Astronomy Observatory, Charlottesville, VA 22903, USA}}

\footnotetext{\dag~Electronic Supplementary Information (ESI) available: Model inputs, raw abundances and calculated sensitivities, and data analysis functions with example Jupyter notebook. See DOI: https://doi.org/10.5281/zenodo.13257329.}



\section{Introduction}
To date, over 300 molecules and ions have been detected in the interstellar medium (ISM), with the number rapidly growing \cite{mcguire_2021_2022}. Although such detections confirm the presence of individual molecular species and provide information about their abundances, the chemical processes surrounding these species often remain a mystery. Astrochemical kinetic models are thus typically relied upon to gain information about the underlying chemical pathways \cite{herbst_formation_1973}. Such models simulate the chemical evolution of an interstellar region given a number of input parameters regarding physical conditions and chemical processes \cite{garrod_three-phase_2013, ruaud_gas_2016, holdship_uclchem_2017, millar_umist_2024}. In larger models this can easily lead to thousands of parameters, all of which must be well-constrained for a robust and reliable model \cite{wakelam_2014_2015}. However, the accuracy of these parameters depends on the available observational, experimental, and theoretical data, which is limited \cite{hebrard_how_2009, wakelam_sensitivity_2010}. In particular, astrochemical models are often highly sensitive to the rate coefficients of gas-phase reactions \cite{hebrard_how_2009, agundez_chemistry_2013}. 

The experimental measurement of rate coefficients in general is a time-consuming and difficult task that is further exacerbated by challenges in measuring these values at low temperature in the gas phase. Techniques developed to meet this challenge include ion cyclotron resonance \cite{mcmahon_versatile_1972}, flowing afterglow \cite{dunkin_ionmolecule_1968}, heavy ion storage rings \cite{geppert_dissociative_2004}, and CRESU \cite{rowe_study_1998}. This last technique is able to study ion-neutral and neutral-neutral reactions at temperatures as low as 10 K \cite{cooke_experimental_2019, hays_collisional_2022}, but requires a very large pumping capacity \cite{rowe_study_1998}. Theoretical treatments also exist for calculating reaction rate coefficients, however, highly-accurate ab initio potential energy surfaces are often needed, as the presence of small barriers can significantly affect calculated rate coefficients \cite{clary_fast_1990,smith_reactions_2006,ma_low_2012,cooke_experimental_2019}. This can become further complicated by the presence of open-shell radicals, which participate in a large portion of reactions relevant to the ISM \cite{ma_low_2012,morozov_theoretical_2020}. Furthermore, effects such as tunneling and the formation of van der Waals complexes often become significant at low temperatures \cite{georgievskii_strange_2007,cooke_experimental_2019}. In the absence of low-temperature rate coefficients, either from experiment or theory, values are often estimated by inspection of similar reactions or by extrapolation of high-temperature values \cite{loison_interstellar_2017}. Such estimations are common in astrochemical models but lead to large uncertainties that are often not quantified \cite{hebrard_how_2009}.

Sensitivity analysis techniques are often applied to complex models to gain an understanding of how uncertainties in input parameters affect output parameters \cite{rabitz_sensitivity_1983,saltelli_sensitivity_1999}. The application of sensitivity analysis methods has been extended to a number of research areas including combustion chemistry \cite{gao_uncertainty_2020} and atmospheric chemistry \cite{thompson_effect_1991}, as well as various astrochemical environments such as the atmospheres of Titan \cite{carrasco_uncertainty_2007,hebrard_how_2009,dobrijevic_comparison_2010,hebrard_photochemistry_2013} and Neptune \cite{dobrijevic_effect_1998}, protoplanetary disks \cite{vasyunin_chemistry_2008}, hot cores \cite{wakelam_estimation_2005}, diffuse clouds \cite{vasyunin_influence_2004}, and dark molecular clouds \cite{vasyunin_influence_2004,wakelam_sensitivity_2009,wakelam_reaction_2010,wakelam_sensitivity_2010,heyl_investigating_2023}. These studies have primarily focused on using the global sensitivity analysis approach to study simple and abundant molecular species, both comparing abundance uncertainties to observed values and determining key reactions. Over the past decade or more since these initial studies, modeling codes have become more robust and incorporated a wider variety of processes, such as dust grain chemistry \cite{ruaud_gas_2016}. Likewise, increasing numbers of detected species and studied reactions requires consistent updates to chemical networks.

A large portion of molecular detections have been made in dark clouds - relatively dense regions of space with low temperatures ($\sim$10 K) yet complex chemical inventories. In particular, the first unambiguous detections of polycyclic aromatic hydrocarbons (PAHs) in the ISM have recently been made toward the dark cloud Taurus Molecular Cloud 1 (TMC-1) \cite{mcguire_detection_2021}. PAHs are thought to be abundant throughout the interstellar medium due to ubiquitous spectral features in the infrared and optical/near-infrared \cite{tielens_interstellar_2008}. These detections suggest formation within the cloud through ``bottom-up" processes, yet current astrochemical models dramatically fail to reproduce the observed abundances \cite{mcguire_detection_2021}. In this work, we present the application of two sensitivity analysis techniques to a modern astrochemical model of the dark cloud TMC-1. We investigate the effects of rate coefficient uncertainties on modeled abundances of aromatic species and identify their key reactions, as well as compare the suitability of the two different techniques for this purpose.

\section{Methodology}
\label{sec:Methods}

\begin{figure*}
    \centering
    \includegraphics[width=\textwidth]{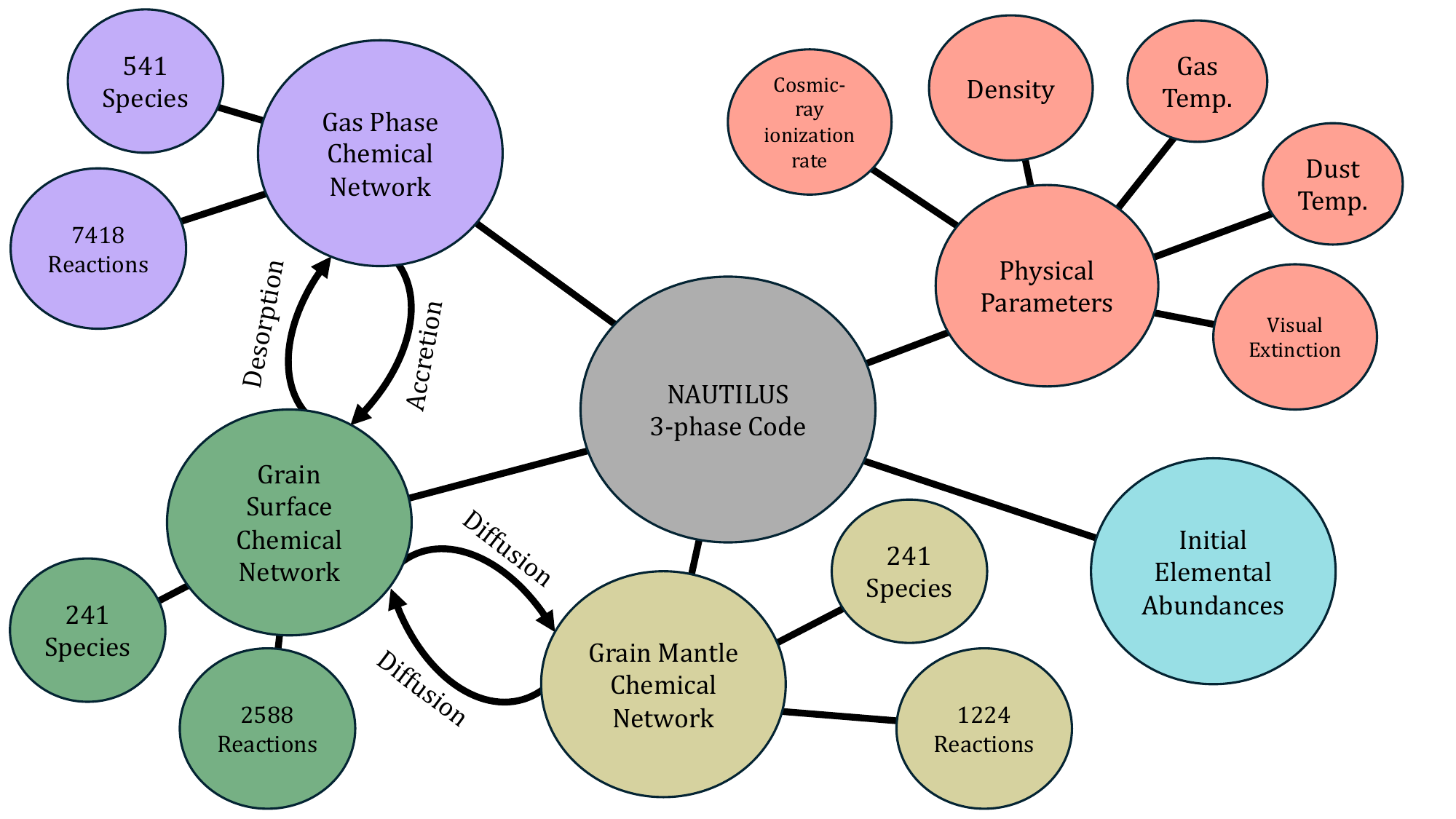}
    \caption{A general schematic of the NAUTILUS 3-phase modeling code. This is a rate-equation model that simulates the chemical evolution of an astronomical region given a chemical network, a number of physical parameters, initial abundances, and various numerical parameters. Species and reactions are separated into gas, grain surface, and grain mantle phases with physical processes allowing for transfer between phases.}
    \label{fig:nautilus}
\end{figure*}

\subsection{Astrochemical Model}

A number of modeling codes and chemical networks have been utilized for astrochemical modeling of TMC-1 and other astronomical objects. \texttt{NAUTILUS} \cite{ruaud_gas_2016}, UCLCHEM \cite{holdship_uclchem_2017}, and RATE22 \cite{millar_umist_2024} are all publicly available rate-equation codes capable of modeling dark clouds with similar general functionality. The former two codes consider physical and chemical processes on interstellar dust grains, while the latter is gas-phase only. We chose the \texttt{NAUTILUS} modeling code as it is a 3-phase code with a number of features based on experimental and theoretical studies. Dust grains are separated into surface and mantle (bulk ice) phases, with separate reactions for the two phases. A general overview of the model can be seen in Figure~\ref{fig:nautilus}.  Accretion of gas-phase chemical species onto grain surfaces, desorption of grain species into the gas-phase, and difussion between grain surface and mantle allow for transfer of species between phases. A number of desorption mechanisms are considered including thermal desorption, cosmic-ray stochastic heating, chemical desorption due to reaction exothermicity, and photodesorption via external UV and cosmic-ray induced photons. Dark cloud chemical networks are often based on the KIDA or UMIST databases. We use gas-phase and grain-phase networks based on kida.uva.2014 \cite{wakelam_2014_2015} and \citet{ruaud_modelling_2015} respectively, which have been used extensively in combination with the \texttt{NAUTILUS} code and its predecessors to study dark cloud chemistry \cite{loison_interstellar_2017, vidal_reservoir_2017, navarro-almaida_gas_2020, doddipatla_low-temperature_2021}. A number of extensions are included based on former works focused on large carbon-chain and aromatic species that have recently been detected in TMC-1 \cite{mcguire_detection_2018, xue_detection_2020, mcguire_early_2020, mccarthy_interstellar_2021, mcguire_detection_2021, loomis_investigation_2021, byrne_astrochemical_2023}. We thus expect our results to be generally applicable to modeling of dark cloud chemistry, while the methods described in the following section are suitable for any kinetic model.

`Typical' dark cloud conditions were assumed, namely a gas density of \SI{2e4}{\per\cubic\centi\metre} \cite{snell_determination_1982}, a kinetic temperature of 10 K for the gas and dust grains \cite{pratap_study_1997}, a cosmic-ray ionization rate, $\zeta$, of \SI{1.3e-17}{\per\second} \cite{spitzer_heating_1968}, and a visual extinction of 10 mag for external UV photons \cite{rodriguez-baras_gas_2021}. Chemical desorption was set to 1\% efficiency, assuming that 1\% of molecules formed on grain surfaces have enough energy to desorb into the gas phase \cite{garrod_non-thermal_2007}. A peak grain temperature of 70 K, lasting for \SI{1e-5}{\second} was used for the cosmic-ray heating mechanism as estimated by \citet{hasegawa_new_1993}. The initial elemental abundances used and their references can be seen in Table~\ref{tab:init_abundances} and represent the low-metal abundances from \cite{graedel_kinetic_1982} with a lower oxygen abundance and a greater sulfur abundance. It has been found that a C/O ratio of 1.1 significantly improves agreement between modeled and observed abundances, particularly for larger hydrocarbons such as unsaturated carbon chains \cite{hincelin_oxygen_2011, xue_detection_2020, loomis_investigation_2021}. Data from a recent survey of sulfur-bearing molecules in molecular clouds suggests a depletion factor of 20 relative to the solar abundance, larger than the previously used value \cite{fuente_gas_2023}. We assume that these elements are all initially in atomic form except for hydrogen, which is primarily molecular. All input files used for the model, including the reaction networks and input parameters, are available in a Zenodo repository: https://zenodo.org/doi/10.5281/zenodo.13257328

\begin{table*}[htb]
    \centering
    \caption{\textbf{Initial atomic/molecular abundances referenced to hydrogen nuclei}}
    \label{tab:init_abundances}
    \centering
    \begin{tabular}{@{} l @{} c c c c c}
    \toprule
    Species & $n/n_H$ & Reference & Species & $n/n_H$ & Reference \\
    \midrule
    \ce{H2} & 0.499 & \citet{ruaud_gas_2016} & \ce{F} & \num{6.68e-9} & \citet{neufeld_chemistry_2005} \\
    \ce{He} & 0.09 & \citet{wakelam_polycyclic_2008} & \ce{Cl} & \num{1.00e-9} & \citet{graedel_kinetic_1982} \\
    \ce{C} & \num{1.70e-4} & \citet{jenkins_unified_2009} & \ce{Si} & \num{8.00e-9} & \citet{graedel_kinetic_1982} \\
    \ce{N} & \num{6.20e-5} & \citet{jenkins_unified_2009} & \ce{Na} & \num{2.00e-9} & \citet{graedel_kinetic_1982}\\
    \ce{O} & \num{1.55e-4} & \citet{xue_detection_2020} & \ce{Mg} & \num{7.00e-9} & \citet{graedel_kinetic_1982} \\
    \ce{S} & {\num{7.50e-7}} & {\citet{fuente_gas_2023}} & \ce{Fe} & \num{3.00e-9} & \citet{graedel_kinetic_1982} \\
    \ce{P} & \num{2.00e-10} & \citet{graedel_kinetic_1982} & & & \\
    \bottomrule
    \end{tabular}
\end{table*}

\subsection{Monte Carlo Analysis}

Previous sensitivity analyses of gas-phase rate coefficients in astrochemical kinetic models, including those of dark molecular clouds, have primarily made use of the Monte Carlo (MC) approach \cite{dobrijevic_effect_1998, vasyunin_influence_2004, wakelam_estimation_2005, wakelam_chemical_2006, carrasco_uncertainty_2007, vasyunin_chemistry_2008, wakelam_sensitivity_2009, hebrard_how_2009, wakelam_sensitivity_2010, wakelam_reaction_2010, hebrard_photochemistry_2013}. In this method, each rate coefficient is selected randomly from an assigned uncertainty distribution, typically a log-normal distribution centered on the nominal value. A large number of iterations are then performed to obtain a statistical spread of abundances \cite{dobrijevic_effect_1998}. Spearman rank correlation coefficients (RCCs) can then be calculated for every rate coefficient-abundance pair. This value is a measure of the monotonic correlation between two variables; essentially how often an increase is observed in both variables simultaneously \cite{spearman_proof_1904}. The reactions with large RCCs can be considered as the reactions whose rate coefficient uncertainties most contribute to uncertainties in modeled abundances, and thus the reactions for which further constraint of rate coefficient would be most beneficial \cite{hebrard_how_2009}. As a global method, the Monte Carlo approach takes into account the entire range of rate coefficient uncertainty and preserves coupling between rate coefficients \cite{rabitz_sensitivity_1983, saltelli_global_2008, dobrijevic_comparison_2010}.

In practice, determining uncertainty distributions for a large number of rate coefficients is a difficult task, particularly when many of them are based on chemical intuition or extrapolation of high-temperature data. Thorough statistical analysis of uncertainties, requiring expertise in the techniques used to obtain these values, has only been performed for a small fraction of reactions present in astrochemical networks \cite{hebrard_how_2009, wakelam_reaction_2010}. As such, we choose to make a blanket assumption and use uncertainties of a factor of 2.0 for every rate coefficient. This factor of 2.0 is the assumed default for most reactions in astrochemical networks and databases \cite{dobrijevic_effect_1998, wakelam_sensitivity_2009, wakelam_reaction_2010}. We also choose to use log-uniform distributions, as they may be more appropriate when uncertainty is not well quantified \cite{gao_uncertainty_2020}. 10,000 iterations of the model were thus performed and RCCs for every species-reaction pair were calculated for 13 time-points between $4.5\times10^4$ and $1.1\times10^6$ years. Monte Carlo analyses with 5000, 15000, and 20000 iterations were also performed to test the convergence.

\subsection{One-At-A-Time Analysis}

In contrast to the global Monte Carlo approach, local sensitivity analysis methods may be used to calculate or estimate the local derivative surrounding the nominal value. As such, these techniques take a calculus-based approach to determining sensitivity rather than a statistical approach \cite{rabitz_sensitivity_1983, saltelli_sensitivity_2005}. The simplest approach is the one-at-a-time (OAT) method, also called the finite differences or brute force method \cite{rabitz_sensitivity_1983, turanyi_applications_1997, dobrijevic_comparison_2010}. This consists of running the model a number of times, where in each iteration one rate coefficient is slightly perturbed while the others are kept constant. Sensitivity coefficients are often calculated according to the equation:
\begin{equation}
    S_i(t) = \left|\dfrac{log X_i(t) - log X_0(t)}{log F_i}\right|
    \label{eqa:sens_coefficient}
\end{equation}
Here $X_i(t)$ is one of the 20 abundances at time $t$ from the $ith$ factor of change, $X_0(t)$ is the nominal abundance of the given species at time $t$, $F_i$ is the $ith$ factor of change, and $S_i(t)$ is the resulting sensitivity coefficient \cite{dobrijevic_comparison_2010}. While this approach has a number of benefits such as being simple and easy to interpret and providing quantitative information, it is often not applicable to nonlinear models where the local derivative is not representative of the greater uncertainty range.

The application of local sensitivity analysis to astrochemical models is somewhat limited. An OAT approach has been applied to hot-core \cite{wakelam_estimation_2005} and dark cloud \cite{wakelam_sensitivity_2009} networks where key reactions were identified based on sensitivity coefficients using either individual rate coefficient uncertainties or an assumed factor of two as the factor of perturbation. \citet{dobrijevic_comparison_2010} applied a number of sensitivity analysis techniques to a photochemical network of Titan's atmosphere, finding that key reactions identified by OAT and MC techniques can differ significantly when there is non-linearity in the model as a result of large uncertainties in rate coefficients. However, this problem may be bypassed through the use of multiple change factors sampled over a larger range of uncertainty, a variation on the Morris method \cite{morris_factorial_1991, saltelli_global_2008}. We perform an OAT analysis using 20 factors of change from 0.5 to 2.0 that are equally spaced in log-scale, assuming a factor of two uncertainty in all rate coefficients as with the MC analysis. For each of the 7418 reactions in our gas-phase network, 20 iterations of the model are performed. In each iteration, the rate coefficient of that reaction is multiplied by one of the 20 change factors, and the resulting modeled abundances of all species are collected at the same 13 time points as the MC analysis. Combined with the nominal, unmodified network this results in 21 abundances for every combination of species, reaction, and time. Three sensitivity metrics are calculated from these sets of 21 abundances. The first, hereafter referred to as SC, uses Equation~\ref{eqa:sens_coefficient} to determine an absolute sensitivity coefficient for each factor of change and averages them for a mean sensitivity coefficient. The other two are based on standard deviation, either dividing by the mean abundance to obtain a relative standard deviation (RSD) or calculating standard deviation using log-scale abundances (log SD). Directionality of the correlation is then included by fitting a linear regression to the abundances as a function of change factor and multiplying the three sensitivity metrics by the sign of the slope.

\section{Results \& Discussion}
\label{sec:Results}

\subsection{Overall Network}

The sensitivities of the overall network to individual reactions were calculated as the average absolute sensitivities of all species in the network to each reaction. This procedure was repeated for the four aforementioned metrics and for three time points of interest. Reactions were then ranked by decreasing average sensitivity to determine key reactions. The chosen time points for comparison are 106500, 314400, and 541100 years, and will be referred to as $1\times10^5$, $3\times10^5$, and $5\times10^5$ years respectively for simplicity. Astrochemical models often agree that the age of TMC-1 seems to be between $10^5$ and $10^6$ years based on comparison to observations, however the exact point within this range is not known. These selected time points thus represent a range of suggested chemical ages, with the later time of $5\times10^5$ generally providing the best agreement between our model and observations of larger hydrocarbons \cite{xue_detection_2020, loomis_investigation_2021, byrne_astrochemical_2023}. Hereafter we refer to the rank of a reaction as its importance, with the highest rank reaction being the most important. For the purpose of comparison, we generally show the top 10 reactions, denoted as key reactions, unless otherwise specified. This is a somewhat arbitrary cutoff as there is no strict definition on what constitutes a ``key'' reaction but is adequate for identifying predominant reactions and comparing these reactions across different metrics and time points. Full reactions lists sorted by average sensitivity metric can be found in the Zenodo repository: https://zenodo.org/doi/10.5281/zenodo.13257328

\begin{figure}[htb]
    \centering
    \includegraphics[width=\columnwidth]{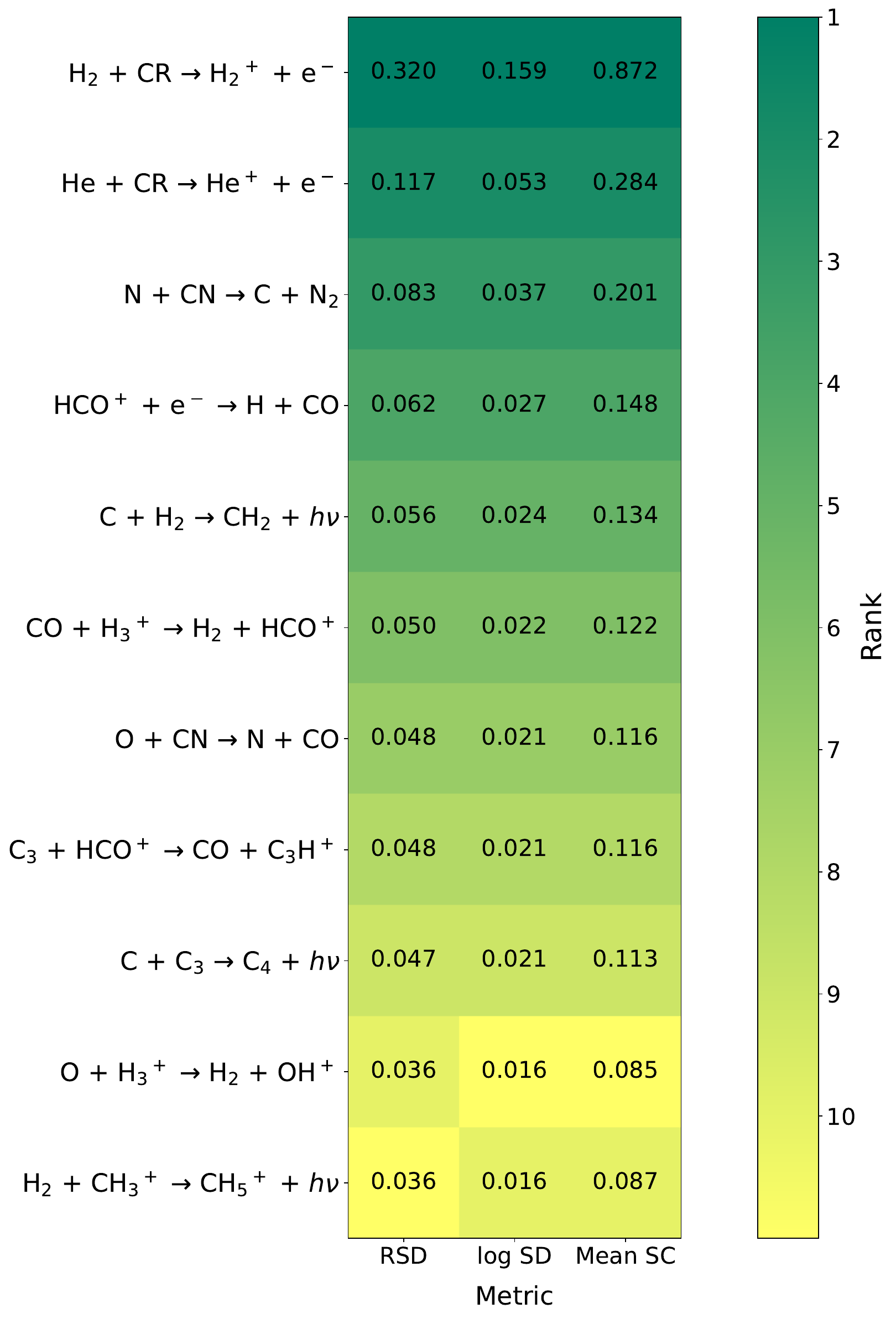}
    \caption{Heat map of key reactions according to the three OAT metrics at $5\times10^5$ years for the overall network. All reaction within the top 10 with any of the three metrics are shown. The shading of each cell corresponds to the reaction's rank, with the exact sensitivity metric also given. Reactions ranked \#11 or lower are given the lightest shading.}
    \label{fig:OAT-overall}
\end{figure}

\begin{figure*}[ht]
    \centering
    \includegraphics[width=\textwidth]{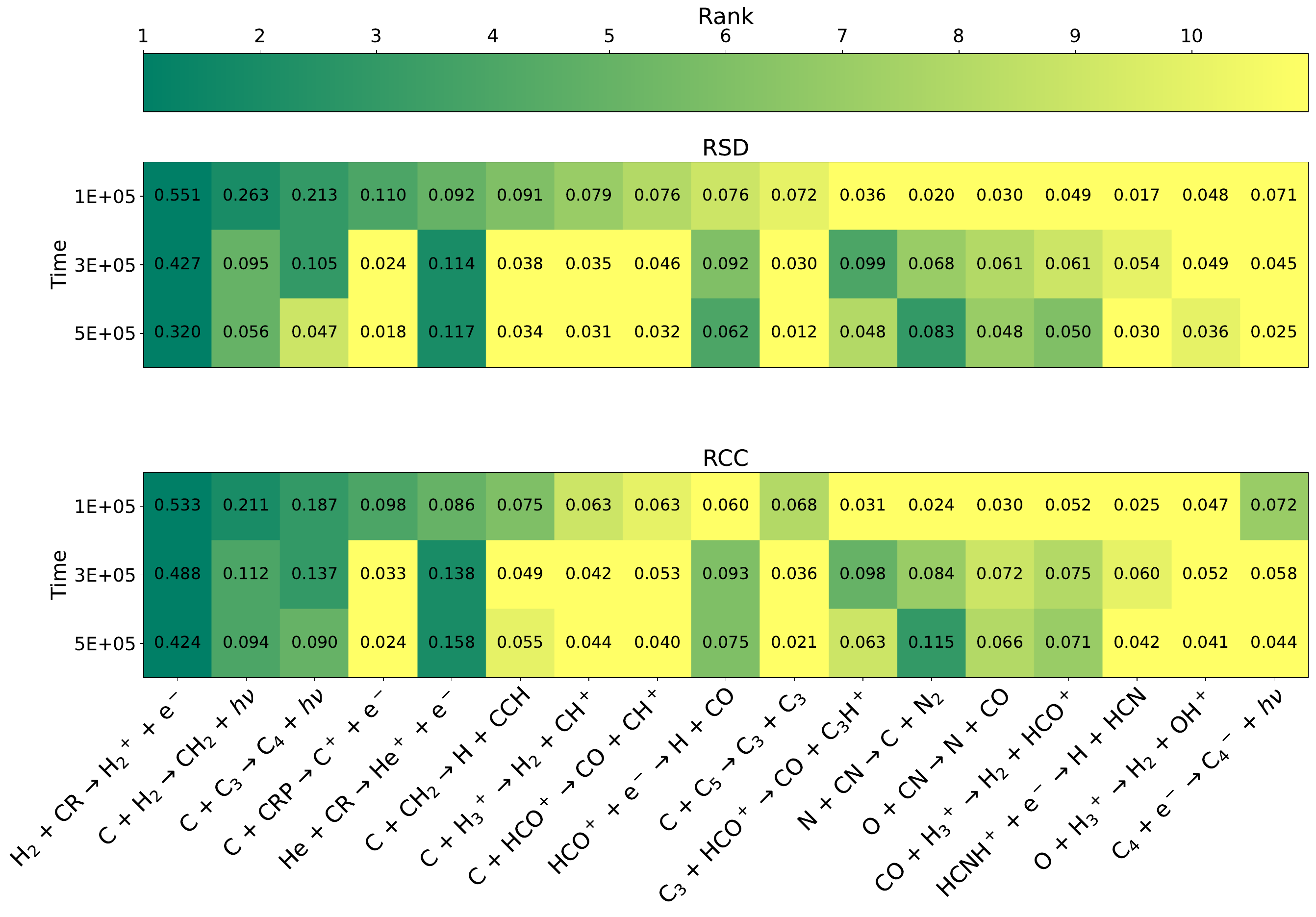}
    \caption{Heat map of key reactions according to the RSD metric (top) and RCC metric (bottom) for the overall network. All reaction within the top 10 at any of the 3 time points with either metric are shown. The shading of each cell corresponds to the reaction's rank, with the exact sensitivity metric also given. Reactions ranked \#11 or lower are given the lightest shading.}
    \label{fig:overall_hmap}
\end{figure*}

Figure~\ref{fig:OAT-overall} shows the top 10 reactions for the overall network at $5\times10^5$ years for all three OAT metrics, ordered by average relative standard deviation. All three metrics agree very well in terms of the reactions identified as key reactions and the relative ordering, with only minor differences in ranking for the bottom two reactions. Additionally, the three metrics agree well in terms of relative magnitude for the sensitivity metrics. The cosmic-ray ionization of molecular hydrogen,
\begin{equation}
    \label{eqa:H2CRIR}
    \ce{H2 + CR -> H2+ + e-},
\end{equation}
is consistently the most important reaction by a significant margin, with the average sensitivity of the next most important reaction only being $\sim$a third of this value. This reaction is the limiting step to formation of \ce{H3+}, which subsequently donates a proton to C or O to initiate hydrocarbon and oxygen chemistry respectively. This reaction also has a strong positive influence on atomic hydrogen abundance, likely due to reaction \ce{H2 + H2+ -> H + H3+} reaction. As a result, the abundances of many protonated and saturated species increase significantly as the cosmic ray ionization rate of \ce{H2} increases. It is important to note that the uncertainty in this rate, along with all other chemical processes involving cosmic rays, can be thought of as a combination of physical uncertainty in the cosmic-ray flux through this region (established through the input parameter $\zeta$) and uncertainty in the efficiency of the chemical process, including uncertainty in photodissociation/photoionization cross-section \cite{vasyunin_chemistry_2008, heays_photodissociation_2017}. As we are altering the rates of individual cosmic-ray processes and not the parameter $\zeta$, we are investigating the latter source of uncertainty. In the case of Reaction~\ref{eqa:H2CRIR}, this arises primarily due to uncertainty in the rate of ionization via secondary electrons, which depends on the gas composition \cite{glassgold_heating_1973,dalgarno_electron_1999}.

The top 10 reactions for the overall network at $1\times10^5$, $3\times10^5$, and $5\times10^5$ years according to the RSD and RCC metrics are shown in Figure~\ref{fig:overall_hmap}. Comparison of key reactions at different time points provides information about how the chemistry evolves as a function of time. The key reactions identified agree well between the OAT and MC methods at all times despite using different definitions of sensitivity. Thus the reaction rates that are correlated with the abundances of many species are generally the same rates for which changes lead to large deviations in abundances of many species. Previous comparison of a number of sensitivity analysis techniques on a small model of Titan's atmosphere found that OAT and MC methods agree well when uncertainties are small, but as uncertainties grow large the agreement can worsen \cite{dobrijevic_comparison_2010}. The authors attribute this to nonlinear behavior that cannot be observed with local methods, and thus suggest that global techniques such as MC should be preferred for uncertainty analysis. The strong agreement between MC and OAT over a larger range of variation may indicate that the use of multiple change factors is able to capture nonlinear behavior, which will be discussed later. It also suggests that coupling between rate coefficients, which is preserved through the MC method but not the OAT method, is not very significant.

Over this range of times, Reaction~\ref{eqa:H2CRIR} is consistently the most important reaction by a large margin. At $1\times10^5$ years the remaining key reactions are dominated by reactions involving atomic carbon. As identified in a previous sensitivity analysis by \citet{wakelam_sensitivity_2009}, many of these are ``early-stage'' reactions at the beginning of hydrocarbon growth. The authors also highlight the reaction of atomic carbon with \ce{C3}, attributing the influence of this reaction to the catalytic conversion of C and O into CO via the reaction
\begin{equation}
    \label{eqa:O+C4}
    \ce{O + C4 -> CO + C3}.
\end{equation}
As time progresses the magnitude of sensitivities generally decreases, suggesting that the network is approaching a steady-state solution and thus becoming less susceptible to individual changes in reaction rates. In particular the importance of reactions involving atomic carbon drastically decrease as it is converted into more complex molecules, although its reactions with \ce{H2} and \ce{C3} remain important. A number of reactions involving formation and destruction of \ce{HCO+} become important at later time, likely due to its role in protonating neutral species such as carbon-chain species. The protonation of \ce{C3} specifically is a critical step for forming \ce{C3H$_n$} species as well as larger hydrocarbons. The cosmic-ray ionization of helium and the reaction
\begin{equation}
    \label{eqa:N+CN}
    \ce{N + CN -> C + N2}
\end{equation}
notably increase in sensitivity as a function of time. The former process is the limiting step to \ce{C+} and atomic oxygen formation via the reaction
\begin{equation}
    \label{eqa:He++CO}
    \ce{He+ + CO -> O + C+}.
\end{equation}
These species react with \ce{C$_m$H} and \ce{C$_m$H2} hydrocarbons that are abundant at later times. Silicon-containing molecules also appear to be highly sensitive to the helium cosmic-ray ionization rate, possibly due to the enhanced production of organo-silicon compounds. Reaction~\ref{eqa:N+CN} has been previously investigated as a major mechanism for partitioning elemental nitrogen into \ce{N2}. We also find that a number of large, more-saturated hydrocarbons such as \ce{C10H8}, \ce{C6H6}, and \ce{C9H5+} are highly sensitive to this rate. Additionally, there is a strong negative correlation between this rate and the abundance of atomic carbon at later times, despite this being a significant formation mechanism for atomic carbon. This seems to suggest that this reaction has a role in the partitioning of elemental carbon into its reactive, atomic form that is a building block for large hydrocarbons. 

For several of these reactions that have been studied experimentally and theoretically, \citet{wakelam_reaction_2010} compiled this data into datasheets along with recommended rate coefficients and uncertainties. The rate coefficient for the reaction
\begin{equation}
    \ce{O + CN -> N + CO}
\end{equation}
has been measured a number of times at room temperature, however the temperature dependence of this reaction is unknown. To our knowledge, the reactions
\begin{align}
    \ce{C + H2 &-> CH2 + {h\nu}} \\
    \ce{C + CH2 &-> H + CCH}
\end{align}
have not been studied experimentally or theoretically and presently use estimated values. Likewise, the majority of the presented ion-neutral reactions do not have experimentally determined constants, although these rates can often be well-predicted by capture theories \cite{su_parametrization_1982}. Further constraining the rates of these critical reactions will be significantly beneficial toward creating more robust models of molecular clouds.

\subsection{Cyanonaphthalene}

\begin{figure}
    \centering
    \includegraphics[width=\columnwidth]{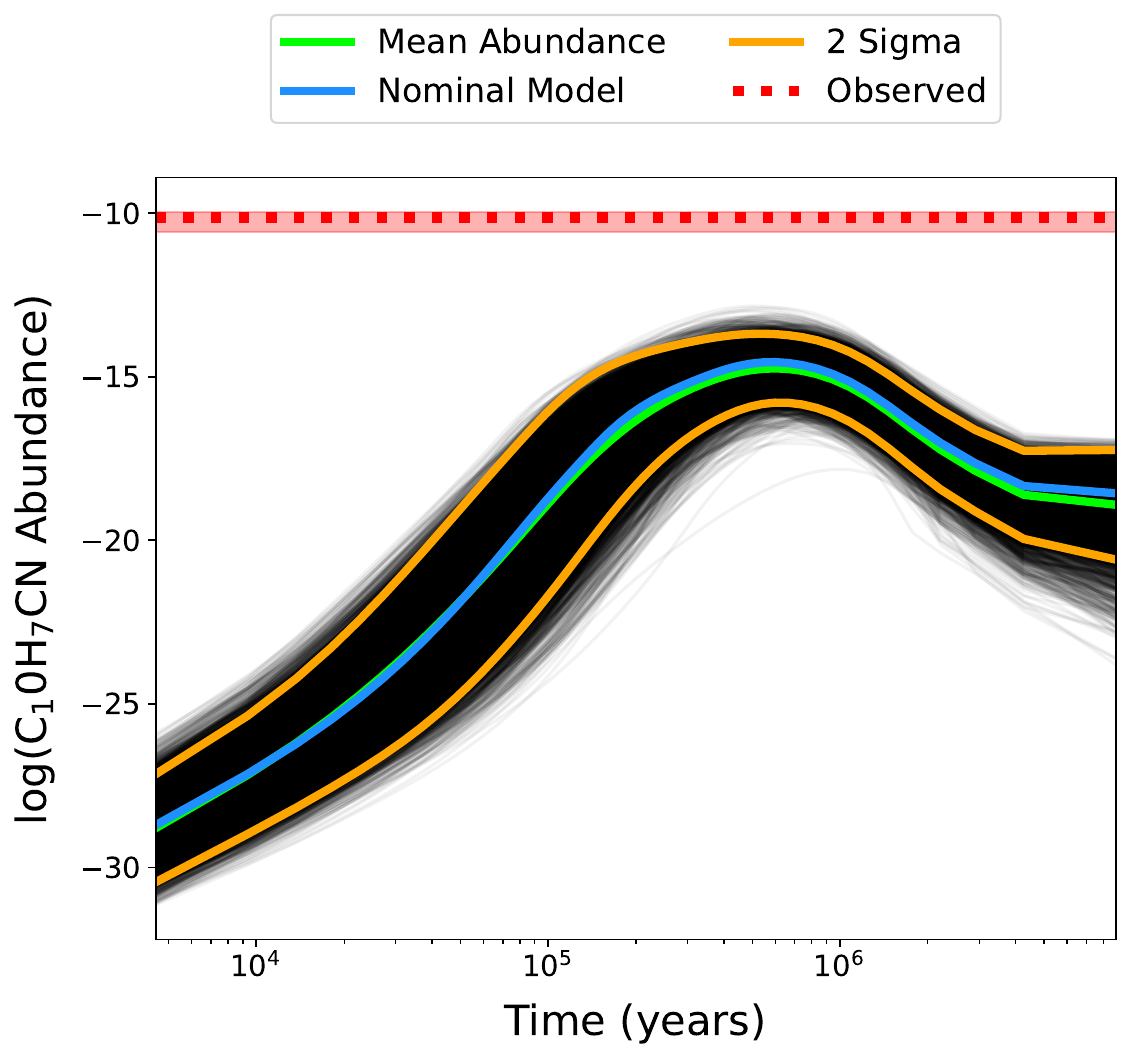}
    \caption{Abundances of \ce{C10H7CN} as a function of time for each of the 10,000 MC iterations. The mean abundance curve is shown in green, along with the nominal model (no changes to rate coefficients) in blue and the $2\sigma$ distribution in yellow. The dotted red line indicates the observed TMC-1 abundance, with the shaded region signifying the $1\sigma$ error.}
    \label{fig:C10H7CN-MC}
\end{figure}

\begin{table*}[htb]
    \centering
    \caption{\textbf{Statistical Analysis of MC Abundances (10,000 iterations) for Select Species at $5\times10^5$ Years}}
    \label{tab:cyanonaphthalene-MC-stat}
    \centering
    \begin{tabular}{l c c c c c c}
    \toprule
    Species & Mean Abundance & +2 $\sigma$ & -2 $\sigma$ & Max Abundance & Min Abundance & Observed Abundance \\
    \midrule
    \ce{C10H7CN} & \num{1.73e-15} & \num{2.05e-14} & \num{1.46e-16} & \num{1.43e-13} & \num{3.04e-19} & \num{7.35e-11} \\
    \ce{C10H8} & \num{5.96e-14} & \num{7.54e-13} & \num{4.72e-15} & \num{5.78e-12} & \num{8.52e-17} & --- \\
    \ce{C6H5} & \num{1.04e-10} & \num{5.76e-10} & \num{1.88e-11} & \num{2.08e-09} & \num{9.62e-13} & --- \\
    \ce{CH2CCH} & \num{1.39e-10} & \num{3.78e-10} & \num{5.08e-11} & \num{1.03e-9} & \num{2.03e-11} & \num{1.00e-8} \\
    \bottomrule
    \end{tabular}
\end{table*}

While the determination of key reactions for the overall network provides information about which processes are generally important due to their effects on a large number of species, a more useful application of sensitivity analysis results may be to study the key reactions for certain species of interest. As the number of detected species grow, more and more of which are larger complex organic molecules, the challenge of modeling these species and understanding the critical pathways to their formation grows as well. Here in particular, we turn our interest toward the PAHs 1- and 2-cyanonaphthalene, which we will hereafter refer to as \ce{C10H7CN}. These species, along with indene \cite{burkhardt_discovery_2021} and cyanoindene \cite{sita_discovery_2022}, are the only PAHs unambiguously detected in the interstellar medium \cite{mcguire_detection_2021}. Previous modeling work has been unable to reproduce the observed abundances by a large margin. However, it is unknown whether the discrepancy arises from missing pathways, underestimated rate coefficients, or both, as well as where in the network these missing pathways or underestimated rates may be. 

In Figure~\ref{fig:C10H7CN-MC}, the time-dependent abundances of \ce{C10H7CN} for all 10,000 MC iterations are plotted, while Table~\ref{tab:cyanonaphthalene-MC-stat} contains the corresponding statistical information for \ce{C10H7CN} and some precursor species. As molecular complexity increases the spread of abundances does as well. This is consistent with the increased rate coefficient uncertainty propagated as the amount of reactions necessary to form these species increases. It is evident that a factor of 2 uncertainty in all reactions is not enough to account for the observed abundance of \ce{C10H7CN}, as the MC iteration with the maximum abundance model still under-produces this species by almost two and a half orders of magnitude compared to its TMC-1 observed value. This is not entirely surprising given the large discrepancies between the nominal model and observations; however it does indicate that much larger uncertainties and/or missing reactions are responsible. We thus turn our attention to the reactions identified as key processes for \ce{C10H7CN} by sensitivity analysis to understand critical processes and areas of the network for future study.

\begin{figure}[htb]
    \centering
    \includegraphics[width=\columnwidth]{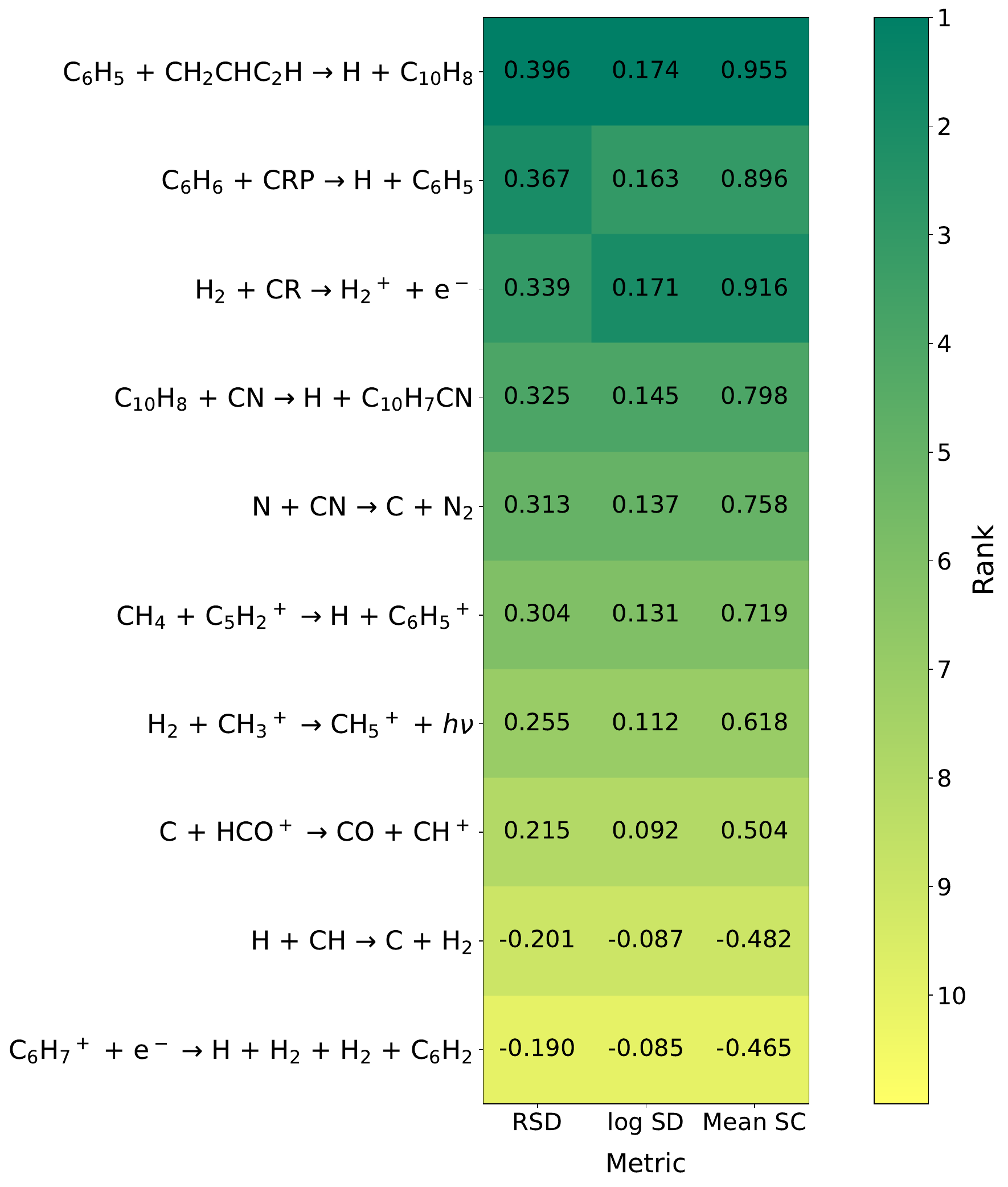}
    \caption{Heat map of key reactions according to the three OAT metrics at $5\times10^5$ years for \ce{C10H7CN}. All reaction within the top 10 with any of the three metrics are shown. The shading of each cell corresponds to the reaction's rank, with the exact sensitivity metric also given. Reactions ranked \#11 or lower are given the lightest shading.}
    \label{fig:OAT-C10H7CN}
\end{figure}

\begin{figure*}[ht]
    \centering
    \includegraphics[width=\textwidth]{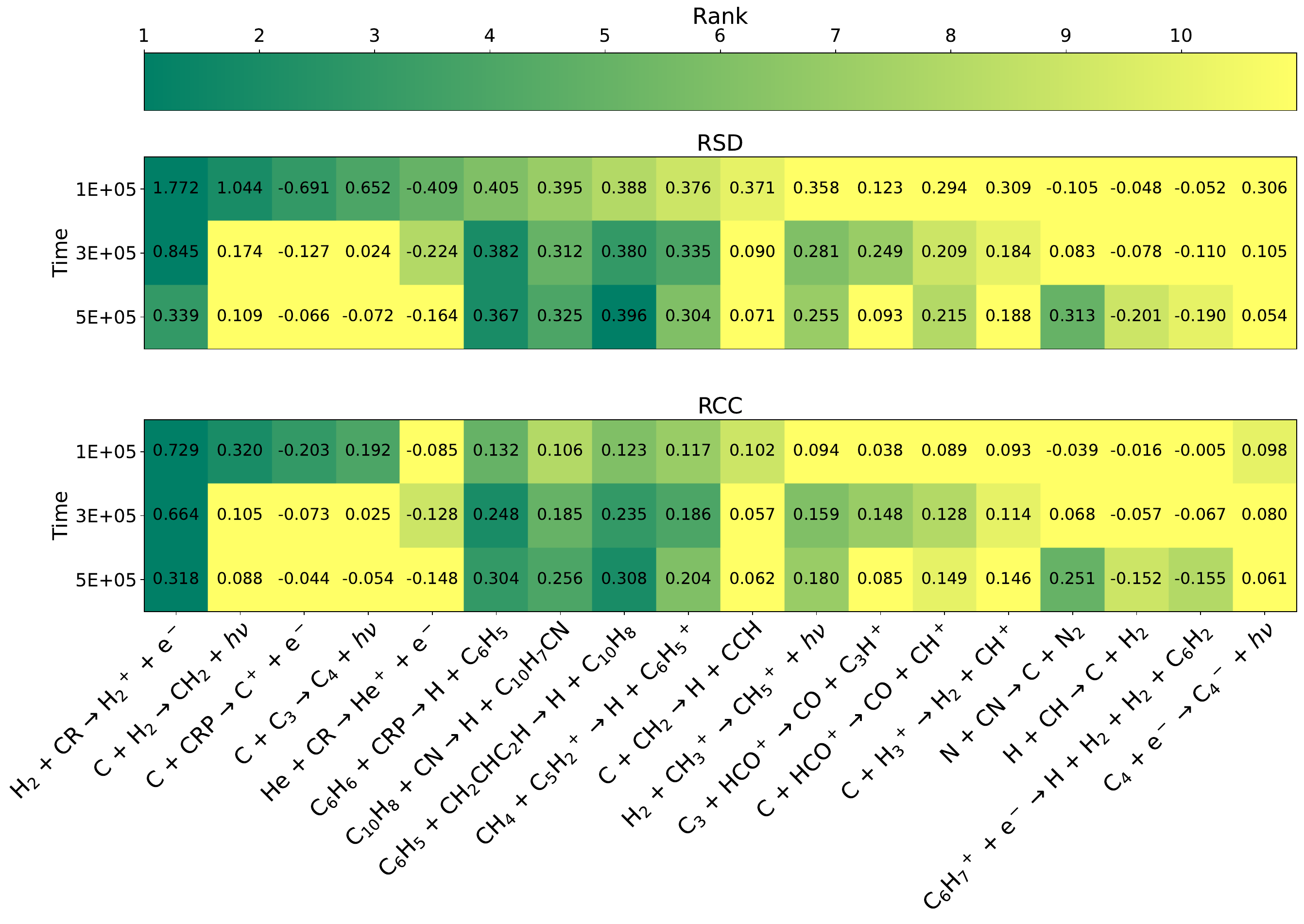}
    \caption{Heat map of key reactions according to the RSD metric (top) and RCC metric (bottom) for \ce{C10H7CN}. All reaction within the top 10 at any of the 3 time points with either metric are shown. The shading of each cell corresponds to the reaction's rank, with the exact sensitivity metric also given. Reactions ranked \#11 or lower are given the lightest shading.}
    \label{fig:C10H7CN_hmap}
\end{figure*}

\subsubsection{Key Reactions and their Rate Coefficients}

The top 10 reactions ranked by sensitivity of \ce{C10H7CN} according to the three OAT metrics at $5\times10^5$ years can be seen in Figure~\ref{fig:OAT-C10H7CN}. Figure~\ref{fig:C10H7CN_hmap} displays these reactions at three select time points according to relative standard deviation and rank correlation coefficients. As with the overall network there is strong agreement between all four metrics. We continue to refer to the ranking of a reaction as its importance, in this section specifically the importance to \ce{C10H7CN}. The cosmic-ray ionization of \ce{H2} is by far the most important reaction at $1\times10^5$ and $3\times10^5$ years. At $5\times10^5$ years there are a number of sensitivities of similar magnitude and this is no longer the case.
 
\begin{figure}[t]
    \centering
    \includegraphics[width=\columnwidth]{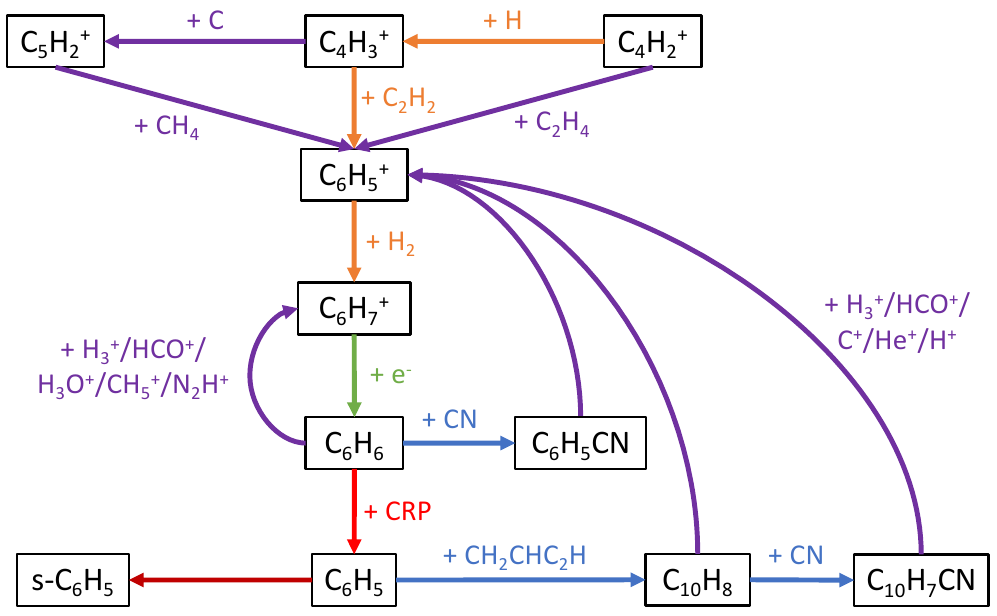}
    \caption{Major chemical reactions leading to formation of aromatic species in our network. A blue arrow indicates a neutral-neutral reaction, an orange arrow radiative association, a green arrow dissociative-recombination, a purple arrow ion-neutral, a red arrow photodissociation, and a brown arrow adsorption onto grains. CRP stands for cosmic-ray induced photons, and the prefix 's-' designates a grain-adsorbed species.}
    \label{fig:aromatic_network}
\end{figure}

These key reactions are a mix of ``early-stage'' hydrocarbon reactions, many of which dominate at early times and are also important to the overall network, and ``late-stage'' aromatic reactions. Reaction network diagrams centered on aromatic and \ce{C3H$_n$} chemistry are showcased in Figures~\ref{fig:aromatic_network} and \ref{fig:propargyl_network} respectively. The reactions of C with \ce{H2}, \ce{C3}, and \ce{CH2} have strong positive correlations with \ce{C10H7CN} abundance as these reactions drive formation of \ce{C3} - \ce{C5} hydrocarbons that are aromatic precursors as can be seen in these figures. Conversely, the cosmic-ray induced photoionization of C and cosmic-ray ionization of He have strong negative correlations with \ce{C10H7CN} abundance as the conversion of C to \ce{C+} and reaction of \ce{C3} with \ce{He+} disrupt the formation of these species. The reactions involving aromatic species primarily consist of ring-formation of the phenyl cation (\ce{C6H5+}) from acyclic hydrocarbons
\begin{equation}
    \label{eqa:CH4+C5H2+}
    \ce{CH4 + C5H2+ -> H + C6H5+},
\end{equation}
the formation of phenyl radical (\ce{C6H5}) and the subsequent formation of naphthalene (\ce{C10H8}) via the HAVA mechanism \cite{parker_low_2012, kaiser_aromatic_2021}
\begin{align}
    \label{eqa:phenyl_rxns1}
    \ce{C6H6 + CRP &-> H + C6H5} \\
    \label{eqa:phenyl_rxns2}
    \ce{C6H5 + CH2CHC2H &-> H + C10H8},
\end{align}
and direct formation of \ce{C10H7CN} itself
\begin{equation}
    \label{eqa:C10H8+CN}
    \ce{C10H8 + CN -> H + C10H7CN}.
\end{equation}
The direct formation of benzene from the dissociative recombination of \ce{C6H7+} does not appear in Figure~\ref{fig:C10H7CN_hmap} but is ranked 11th according to the RSD metric and 13th according to the RCC metric at $5\times10^5$ years. This reaction is fast and dominates the production of benzene; however the relative importance of this process along with the competing product channel to form \ce{C6H2} suggests the branching ratios of these channels appear to have a moderate effect on the abundance of benzene and subsequent species.

At later times, the relative importance of reactions involving atomic carbon generally decreases as was the case for the overall network. Of the reactions that rise in importance, three of them, 
\begin{align}
    \ce{C + HCO+ &-> CO + CH+} \\
    \ce{C + H3+ &-> H2 + CH+} \\
    \ce{H2 + CH3+ &-> CH5+ + {h\nu}}
\end{align}
are part of the pathway from atomic carbon to methane as visible in Figure~\ref{fig:propargyl_network}. As shown, methane is a key precursor to the aromatic cation \ce{C6H5+} as well as \ce{C3} hydrocarbons that participate in the formation of aromatic species. The reaction 
\begin{equation}
    \ce{H + CH -> C + H2}
\end{equation} displays a negative correlation with \ce{C10H7CN} abundance due to depletion of \ce{CH}. In Figure~\ref{fig:propargyl_network}, it can be seen that this radical reacts with isomers of \ce{C3H4} to form vinylacetylene (\ce{CH2CHC2H}), a key component of the HAVA mechanism. Additionally, the relative importance of the four previously mentioned aromatic reactions increases from $1\times10^5$ years to $5\times10^5$ years as simple carbon-containing species are converted into more complex hydrocarbons. Finally, we again note that reaction~\ref{eqa:N+CN} becomes significantly more sensitive with time, becoming the 5th most important reaction to \ce{C10H7CN} at $5\times10^5$ years. As mentioned previously, we suspect this may be due to this reaction's affect on the partitioning of carbon into larger hydrocarbons that are precursors to aromatic species. Notably, the strong positive correlation between \ce{C10H7CN} abundance and the rate of this reaction further suggests that this is the predominant effect. If removal of CN was as or more important we would expect this correlation, and those of many other CN-containing species, to be negative. 

 Of the four aromatic reactions that the \ce{C10H7CN} abundance is highly sensitive to, none of them have been studied experimentally at low temperature. The rate prefactor for reaction~\ref{eqa:phenyl_rxns1} comes from the high-temperature network of \citet{harada_new_2010}, with the 154 nm branching ratios from \citet{kislov_photodissociation_2004} corresponding the internal UV field expected in dense clouds \cite{heays_photodissociation_2017}. Rate coefficients for cosmic-ray induced photoprocesses are typically calculated as a function of the \ce{H2} cosmic-ray ionization rate, the photodissociation/photoionization cross section of the molecule of interest, the total absorption cross section, and the induced UV spectrum \cite{gredel_cosmic-ray--induced_1989}, although the particular details of this calculation are not provided. More recent calculations by \citet{heays_photodissociation_2017} using updated cross sections yielded general agreement within a factor of 2 as well as some large deviations for specific molecules/atoms. Updated calculations have not yet been performed on \ce{C6H6}. Reaction~\ref{eqa:phenyl_rxns2} presently uses a rate coefficient of \SI{2.5e-10}{\cubic\centi\metre\per\second} and a branching ratio of 100\% according to variational transition state theory and Rice-Ramsperger-Kessel-Marcus (RRKM) calculations at 100 K and zero pressure respectively \cite{parker_low_2012}. Follow-up RRKM calculations at zero pressure and 60 K included radiative stabilization of intermediates, finding that the inclusion of this mechanism could favor the formation of unimolecular products over the \ce{C10H8 + H} channel \cite{mebel_formation_2017}. The direct formation of \ce{C10H7CN} (reaction~\ref{eqa:C10H8+CN}) was estimated to occur with a rate near the collision rate as it is a reaction between an unsaturated hydrocarbon and a neutral radical where the difference in ionization energy of the former and electron affinity of the latter is less than 8.75 eV \cite{smith_temperature-dependence_2006, mcguire_detection_2021}. CRESU measurements at 15 K of the analogous reaction between \ce{C6H6} and CN resulted in a rate coefficient of \SI{5.4e-10}{\cubic\centi\metre\per\second}, indicating that the rate of reaction with \ce{C10H8} could be larger than the currently used value. Finally, reaction~\ref{eqa:CH4+C5H2+} originates from \citet{herbst_gas-phase_1989} with an overall rate coefficient of \SI{1.0e-9}{\cubic\centi\metre\per\second} and 80/20 branching ratios for the H/\ce{H2} elimination channels estimated from \cite{anicich_survey_1986}. In addition to uncertainties in the overall rate, which may be better estimated by capture theory \cite{su_parametrization_1982}, it is unknown whether this is a viable mechanism to formation of the cyclic \ce{C6H5+} isomer. This product channel, which requires breaking and forming a large number of bonds, is likely in competition with formation of acyclic products and its status as the dominant channel is debatable.

\subsubsection{Completeness of the Network}

\begin{figure}[t]
    \centering
    \includegraphics[width=\columnwidth]{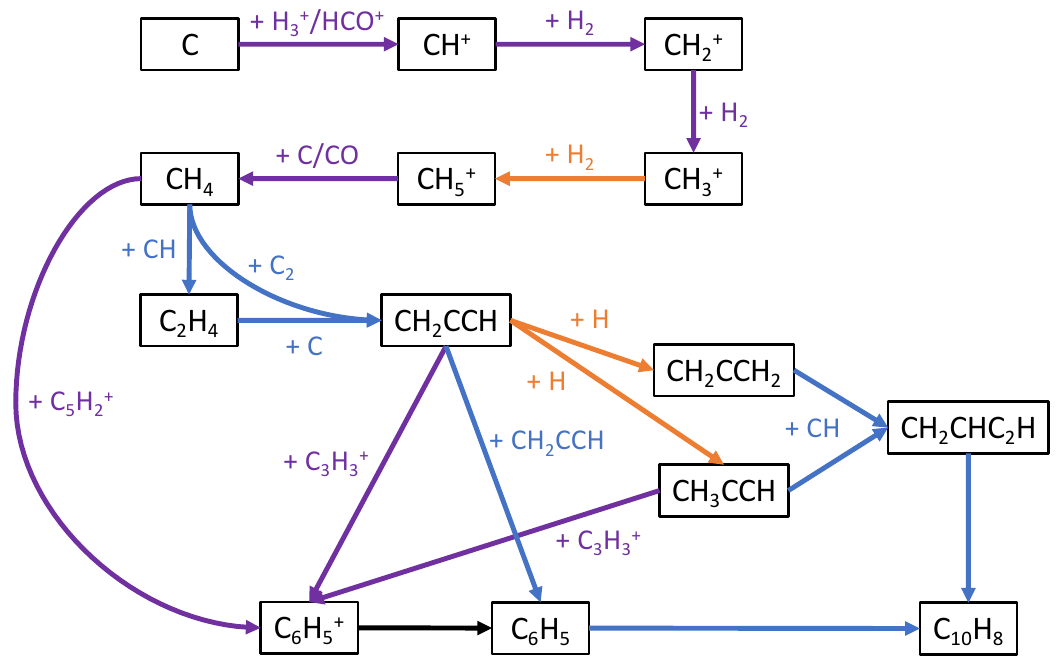}
    \caption{Major chemical reactions leading from atomic carbon to the formation of \ce{CH4}, \ce{C3} hydrocarbons, and six-membered rings. A blue arrow indicates a neutral-neutral reaction, an orange arrow radiative association, and a purple arrow ion-neutral. The black arrow leading from \ce{C6H5+} to \ce{C6H5} is meant to show that pathways exist from the former to the latter, the exact ones being showcased in Figure~\ref{fig:aromatic_network}.}
    \label{fig:propargyl_network}
\end{figure}

Beyond the determination of individual rate coefficients that should be further constrained, the identification of key reactions provides information about areas of the network that are incomplete. As shown in Figures~\ref{fig:aromatic_network} and \ref{fig:propargyl_network}, a number of ring-formation mechanisms to form the first six-membered aromatic ring, primarily \ce{C6H5+}, exist in the network. These include [4 + 2] reactions proposed by \citet{mcewan_new_1999} as well as a number of [3 + 3] reactions suggested by \citet{herbst_gas-phase_1989}. The sensitivity analysis results clearly indicate that reaction~\ref{eqa:CH4+C5H2+}, a [5 + 1] mechanism involving methane, is the most important of these. The next most important ring-forming reaction is 
\begin{equation}
    \ce{C2H2 + C4H3+ -> C6H5+ + {h\nu}}.
    \label{C2H2+C4H3+}
\end{equation}
Despite potential energy surface calculations confirming that this radiative association should be barrierless and efficient \cite{peverati_insights_2016}, it is only rank \#31 in terms of importance to \ce{C10H7CN}. The remaining ring-formation mechanisms of \ce{C6H5+} are significantly less important with ranks of \#161 or lower. Direct formation of \ce{C6H6} \cite{jones_formation_2011}
\begin{equation}
    \ce{CCH + CH2CHCHCH2 -> H + C6H6}
    \label{eqa:CCH+CH2CHCHCH2}
\end{equation}
and of \ce{C6H5} \cite{miller_recombination_2003, zhao_gas-phase_2021}
\begin{equation}
    \ce{CH2CCH + CH2CCH -> C6H5 + H}
    \label{eqa:CH2CCH+CH2CCH}
\end{equation}
are found to be even less important to \ce{C10H7CN} abundance with ranks of \#1116 and \#976 respectively. There is a notable drop-off in reaction significance once neutral hydrocarbons with 3 or more hydrogen atoms (\ce{C2H3}, \ce{C2H4}, \ce{CH2CCH}, \ce{CH3CCH}) or hydrocarbon cations with 4 or more hydrogen atoms (\ce{C4H4+}) are involved. The exception is reaction~\ref{eqa:CH4+C5H2+} involving \ce{CH4}, the simplest saturated hydrocarbon that is efficiently formed from a chain of ion-molecule reactions from \ce{CH+} to \ce{CH5+} followed by proton donation to CO \cite{herbst_synthesis_2017} as shown in Figure~\ref{fig:propargyl_network}. This trend suggests a larger problem in the network regarding limited production of more-saturated hydrocarbons that are needed for ring-formation. A large portion of detected TMC-1 molecules and ions have historically been unsaturated hydrocarbons and their derivatives; however, a number of recent detections have found high abundances of more-saturated species that cannot be reproduced by current astrochemical models \cite{marcelino_discovery_2007, gratier_new_2016, cernicharo_discovery_2021, cernicharo_discovery_2022, cooke_detection_2023}. In particular, the propargyl radical (\ce{CH2CCH}) has been detected in this cloud with a large abundance \citep{agundez_detection_2022}. This species is thought to play a central role in the formation of six-membered aromatic rings from smaller hydrocarbons as indicated in Figure~\ref{fig:propargyl_network}, however the inability of the model to reproduce the observed abundance results in little contribution from \ce{CH2CCH} to ring-formation \citep{byrne_astrochemical_2023}. Likewise, the abundance of \ce{C10H7CN} is not highly sensitive to any reaction directly involving \ce{CH2CCH} nor is the observed abundance of \ce{CH2CCH} reproduced in any of the Monte Carlo iterations. This indicates that even within a factor of two variation the present pathways to \ce{CH2CCH} and other \ce{C3} hydrocarbons from \ce{C1} and \ce{C2} hydrocarbons are inadequate, and other mechanisms should be explored.

Hydrogenation of \ce{C3H$_n$} species via reaction with \ce{H2} has been found to proceed efficiently from \ce{C3+} to \ce{C3H+} followed by radiative association to \ce{C3H3+} based on low temperature ion trap experiments \cite{savic_temperature_2005}. \ce{C3H} and \ce{C3H2} can subsequently be formed by dissociative recombination of \ce{C3H3+} \cite{angelova_branching_2004, mclain_c3h3_2005}. Further hydrogenation to \ce{C3H4+} was found to be slow with rate coefficients on the order of \SI{1.0e-14}{\cubic\centi\metre\per\second}. Calculations on the radiative associations of \ce{C3H3+} and \ce{C3H5+} with \ce{H2} have shown sizeable barriers in the entrance channels \cite{lin_can_2013}. Similarly, selected ion flow tube experiments at room temperature were not able to detect a reaction between \ce{C3H3+} and H, establishing an upper limit of \SI{3.0e-12}{\cubic\centi\metre\per\second} for the rate coefficient \cite{scott_cmhn_1997}. The problem of missing information regarding \ce{C3} and larger hydrocarbons has also been discussed in the context of modeling chemistry in Titan's atmosphere, with a number of radiative association reactions proposed \cite{hebrard_photochemistry_2013}. Rate coefficients for these processes were determined based on a semi-empirical fit to previous calculations using capture rates and densities of states. These reactions are able to efficiently form more-saturated hydrocarbons such as \ce{C3H3} and \ce{C3H6} but have uncertainties in rate coefficients of a factor of 30 or larger. Grain-surface hydrogenation of \ce{C3} up to \ce{C3H8} through consecutive additions of H has been found to be quickly produce \ce{C3H$_n$} molecules on dust grains \cite{hickson_methylacetylene_2016, loison_interstellar_2017}, although additional parameters such as binding energies and efficiency of nonthermal desorption become significant for such processes. An alternative pathway involving radiative attachment of an electron followed by associative detachment with H exists for \ce{C4} and larger carbon chains
\begin{align}
    \ce{C_{m} + e- &-> C_{m}- + {h\nu}} 
    \label{eqa:Cm+e-} \\
    \ce{C_{m}- + H &-> C_{m}H + e-} 
    \label{eqa:Cm-+H} \\
    \ce{C_{m}H + e- &-> C_{m}H- + {h\nu}} 
    \label{eqa:CmH+e-} \\
    \ce{C_{m}H- + H &-> C_{m}H2 + e-}.
    \label{eqa:CmH-+H}
\end{align}
Rate coefficients for reactions~\ref{eqa:Cm+e-} for $4\leq m\leq 9$ and \ref{eqa:CmH+e-} for $2\leq m\leq 8$ have been determined using statistical calculations \cite{terzieva_radiative_2000} and phase-space theory respectively \cite{herbst_calculations_2008}. Room temperature rate coefficients have been measured for reactions~\ref{eqa:Cm-+H} and \ref{eqa:CmH-+H} for a number of carbon-chain lengths using a flowing afterglow-selected ion flow tube \cite{barckholtz_reactions_2001}. This mechanism presently does not extend beyond \ce{C$_m$H2} hydrocarbons, possibly due to competition with dissociative attachment. There is a clear need for future work regarding gas-phase formation of more-saturated hydrocarbons, particularly \ce{C3H$_n$} species with $n\geq3$ that can participate in ring-formation.

It can be seen in Figure~\ref{fig:aromatic_network} that the remaining aromatic reactions that \ce{C10H7CN} is highly sensitive to are the three final steps leading from \ce{C6H6} to \ce{C10H7CN} through \ce{C6H5} and \ce{C10H8}. Given that the two most important reactions to \ce{C10H7CN} involve \ce{C6H5}, this radical appears to be the major limiting species toward formation of PAHs in the network. \ce{C6H5} is predominately formed by cosmic-ray induced photodissociation of \ce{C6H6}, with direct ring-formation from \ce{CH2CCH} recombination inefficient as previously discussed. Alternative H-abstraction mechanisms for \ce{C6H6} such as reaction with H, OH, \ce{CH3}, and \ce{C2H5} all have sizeable barriers. Once \ce{C6H5} is formed, only 3\% or less of it reacts with \ce{CH2CHC2H} to form \ce{C10H8}. Due to a lack of reaction with interstellar cations, the vast majority of \ce{C6H5} adsorbs onto dust grains where it is essentially trapped. Inclusion of additional destruction mechanisms would limit the amount of \ce{C6H5} trapped on dust grains but also the amount available in the gas-phase, thus it may be that mechanisms other than the HAVA mechanism are required to produce PAHs in dark clouds. For example, potential energy surface calculations of the reactions
\begin{align}
    \ce{C6H6 + C4H3+ &-> H + C10H8+} \\
    \ce{C6H6+ + 2C2H &-> C10H8+} \\
    \ce{C6H5+ + 2C2H2 &-> C10H9+}
\end{align}
indicate that \ce{C10H8+} and \ce{C10H9+} may be formed from barrierless ion-molecule processes involving a six-membered ring \cite{bauschlicher_jr_mechanisms_2000, ghesquiere_formation_2013}, although experiments on this latter reaction find a small barrier for the second addition of \ce{C2H2} and the formation of an adduct with a four-membered ring rather than a protonated naphthalene structure \cite{soliman_formation_2012}. Likewise, \ce{C10H7+} can rapidly undergo radiative association with \ce{H2} to form \ce{C10H9+} \cite{le_page_chemical_1997, snow_interstellar_1998}. Neutral \ce{C10H8} may then be formed by charge transfer of \ce{C10H8+} or dissociative recombination of \ce{C10H9+} \cite{bohme_pah_1992, le_page_reactions_1999}. In combustion chemistry, ion-based mechanisms to soot formation involving \ce{C3H3+} have been proposed, with a number of cyclic ions such as \ce{C7H7+}, \ce{C9H7+}, \ce{C9H9+}, \ce{C10H8+}, and \ce{C10H9+} detected as products \cite{baykut_reactions_1986, smyth_ion-molecule_1982, adams_laboratory_2010}. Considering the wide variety of ion-molecule reactions that may be involved in interstellar aromatic chemistry and the large rate coefficients of these processes at low temperatures, further expansion and refinement of the aromatic network with a focus on these processes should be another priority for future work.

The incompleteness of the network can also be seen by comparing the magnitude of \ce{C10H7CN} sensitivities in Figure~\ref{fig:C10H7CN_hmap} to those of the average sensitivities in Figure~\ref{fig:overall_hmap}. The fact that \ce{C10H7CN} is significantly more sensitive to rate coefficients at these times compared to the average species may be a consequence of it being the largest and most chemically `complex' species in the network. A long sequence of reactions including carbon-chain growth, ring-formation, and the HAVA mechanism are required for it to be formed. Once \ce{C10H7CN} is formed, its destruction mechanisms are limited to reaction with abundant cations (\ce{H+}, \ce{He+}, \ce{H3+}, \ce{HCO+}, \ce{C+}, and \ce{H3O+}) or adsorption onto dust grains. The former mechanism is assumed to yield \ce{C6H5+}, which can reform \ce{C10H7CN}. The result is that any increase to reactive flux along the chain of reactions leading to \ce{C10H7CN} accumulates in the abundance of this species, making it a kinetic sink. Other processes such as subsequent HAVA to form larger PAHs, or ethynyl addition are likely competing processes that are not currently included in the network. Extension of the aromatic network should thus encompass such reactions and extend beyond \ce{C10H8} and \ce{C10H7CN}. 

\subsubsection{Model Linearity}

\begin{figure}
    \centering
    \includegraphics[width=\columnwidth]{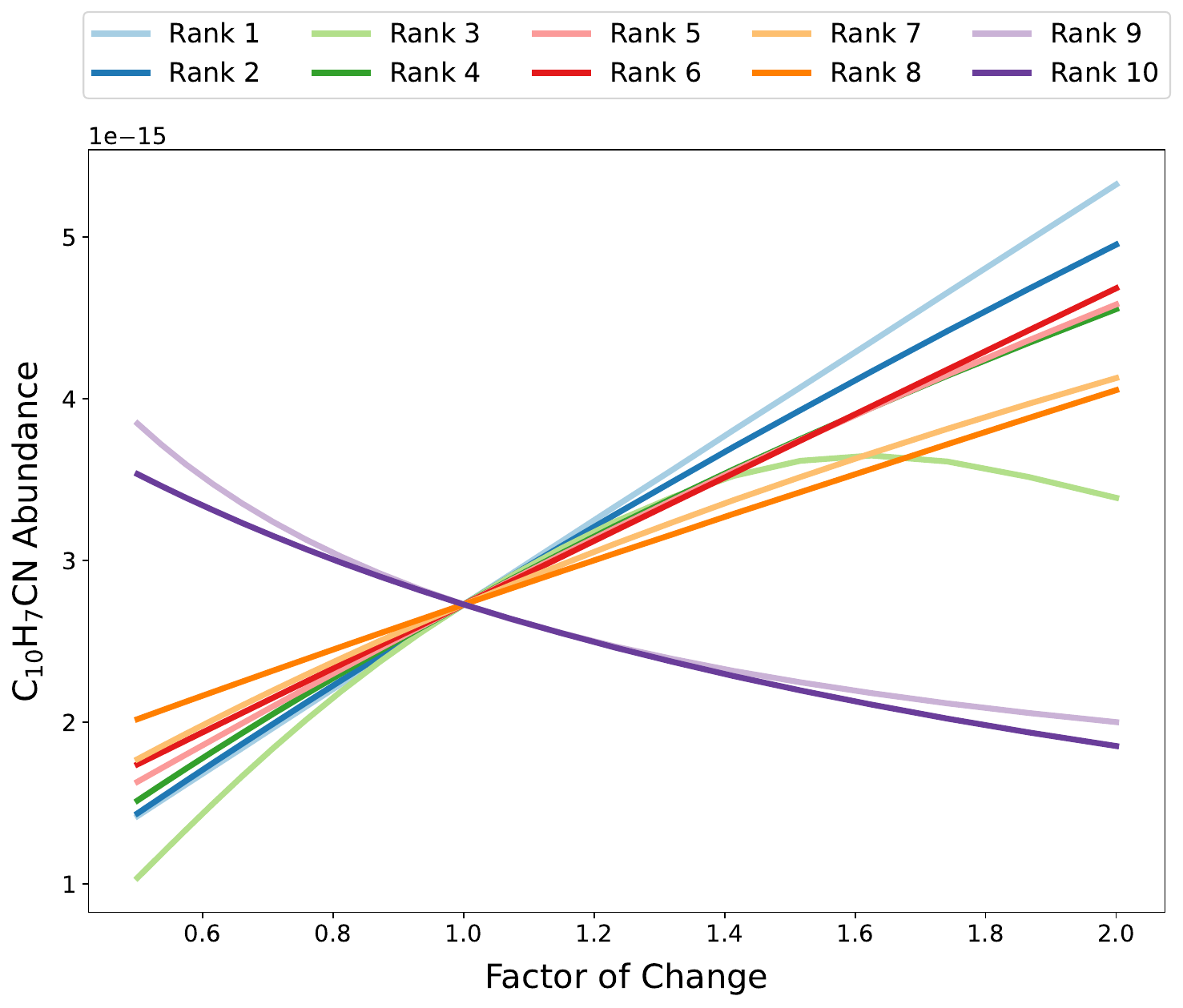}
    \caption{Abundances of \ce{C10H7CN} at $5\times10^5$ years as a function of factor of change in rate constant for the top 10 reactions according to the RSD metric. Each dot represents an abundance when the rate coefficient of the corresponding reaction has been multiplied by some factor, with the color of the dot representing the degree of change. There are 21 factors of change ranging from 0.5 to 2.}
    \label{fig:C10H7CN-OAT}
\end{figure}

In Figure~\ref{fig:C10H7CN-OAT}, the abundance of \ce{C10H7CN} at $5\times10^5$ years as a function of change in rate coefficient from the OAT analysis is shown for its key reactions according to the RSD metric. As can be seen, there is a near-linear relationship between \ce{C10H7CN} abundance and change in rate coefficient for many of these reactions, particularly for the rank 1 reaction \ref{eqa:phenyl_rxns1}. It is clear that the relationship between the rate of reaction rank 3 (which turns out to be Reaction~\ref{eqa:H2CRIR}, the cosmic-ray ionization of hydrogen), is highly nonlinear and non-monotonic. Decreasing this rate coefficient below the nominal value leads to a strong exponential decrease in \ce{C10H7CN} abundance. Conversely, increasing this rate coefficient above the nominal value results in a stationary point of peak \ce{C10H7CN} abundance corresponding to a change factor of $\sim1.6$, after which \ce{C10H7CN} abundance begins decreasing. This is likely due to competition between hydrocarbon growth and destruction, both of which proceed via ion-neutral reactions initiated by \ce{H3+} formation. A greater rate of \ce{H2} cosmic-ray ionization allows complex hydrocarbons to be formed more quickly and to a greater degree, after which destruction via reaction with ions such as \ce{H3+}, \ce{HCO+}, and \ce{N2H+} take over. The ability of our OAT analysis to capture this relationships and assign appropriately large sensitivity coefficients shows that the use of multiple change factors enables the acquisition of quantitative, reaction-specific sensitivities over a wide range of variations.

\section{Conclusion}
\label{sec:Conclusion}
We have performed Monte Carlo and one-at-a-time sensitivity analyses of a 3-phase dense cloud astrochemical model containing a number of complex organic molecules, including aromatic species. By using multiple factors of change for the one-at-a-time analysis, a larger parameter space can be explored and nonlinear relationships can be discovered. The resulting sensitivities agree well with those determined statistically via the Monte Carlo approach. We find that reactions involved in the initiation of interstellar chemistry through the formation of small reactive species, particularly those responsible for carbon-chain growth, have the greatest average effect on modeled abundances. Between $1\times10^5$ and $5\times10^5$ years, the likely age range of TMC-1, the absolute sensitivities and their relative orderings can change significantly. A general decrease in sensitivity metrics with time suggests the approach of a steady-state solution, with reactions involving atomic carbon becoming relatively less important at later times due to increasing complexity of carbon-containing species. Several of these key reactions still use estimated rate coefficients with large uncertainties, and as such more accurate values would significantly benefit the model. Likewise, the use of sensitivity analysis techniques allows us to reveal interesting behavior in the model even for reactions with well-constrained rate coefficients, such as the nonlinear relationship between the rate of reaction of N with CN and the abundance of atomic carbon.

Application of these results to the PAH cyanonaphthalene reveals that it is notably sensitive to four particular aromatic reactions: 1) the ring-formation of \ce{C6H5+} from \ce{C5H2+} and \ce{CH4}, 2) the photo-dissociation of benzene to \ce{C6H5} via the cosmic-ray induced UV field and 3) its subsequent reaction with \ce{CH2CHC2H} to form naphthalene, and 4) direct formation of cyanonaphthalene via a CN-addition H-elimination mechanism. Consequently the major bottlenecks to \ce{C10H7CN} formation appear to be formation of both the first and second aromatic ring. A lack of experimental data and robust calculations casts significant uncertainty on the rates of these reactions. It is also unclear whether the reaction between \ce{CH4} and \ce{C5H2+} is a viable pathway to \ce{C6H5+}, as well as to what degree radiative stabilization competes with the bimolecular product channel for the formation of \ce{C10H8} from the HAVA mechanism. Due to these uncertainties these reactions appear to be good candidates for future experimental and theoretical kinetic and mechanistic studies. Likewise, the high sensitivity of \ce{C10H7CN} to these reactions suggests that the network may be incomplete in these areas. Alternate pathways to the first six-membered ring are hindered by inefficiencies in formation of more-saturated \ce{C3} and \ce{C4} hydrocarbons, possibly due to missing hydrogenation mechanisms such as radiative association with \ce{H2} or dissociative attachment. Additional ion-neutral routes to form the second aromatic ring may be feasible and rapid under TMC-1 conditions considering their general lack of activation barriers and rate coefficients near the collision rate. The identification of key reactions and incomplete regions of chemistry provides valuable insight for future work on improving kinetic models of dark molecular clouds. Additionally, these methods are broadly applicable to other areas of astrochemical modeling, as well as to fields such as combustion chemistry and atmospheric chemistry where kinetic models are also extensively used.

\section*{Author Contributions}
ANB: Conceptualization, methodology, investigation, formal analysis, data curation, visualization, and writing - original draft. CX: Methodology and software. TVV: Conceptualization, supervision, and project administration. BAM: Conceptualization, resources, supervision, and project administration. All authors contributed to writing - review and editing.

\section*{Conflicts of interest}
There are no conflicts to declare.

\section*{Acknowledgements}
We thank Dr. Valentine Wakelam for use of the \texttt{NAUTILUS} v1.1 code. ANB acknowledges support from the National Science Foundation Grant Number 2141064. B. A. M. and C. X. gratefully acknowledge support of National Science Foundation grant AST-2205126.



\balance


\bibliography{Sensitivity} 
\bibliographystyle{rsc} 

\clearpage

\onecolumn
\begin{appendices}

\section{One-at-a-time Analysis with Order of Magnitude Variation}

\begin{figure*}[hb]
    \centering
    \includegraphics[width=.37\textwidth]{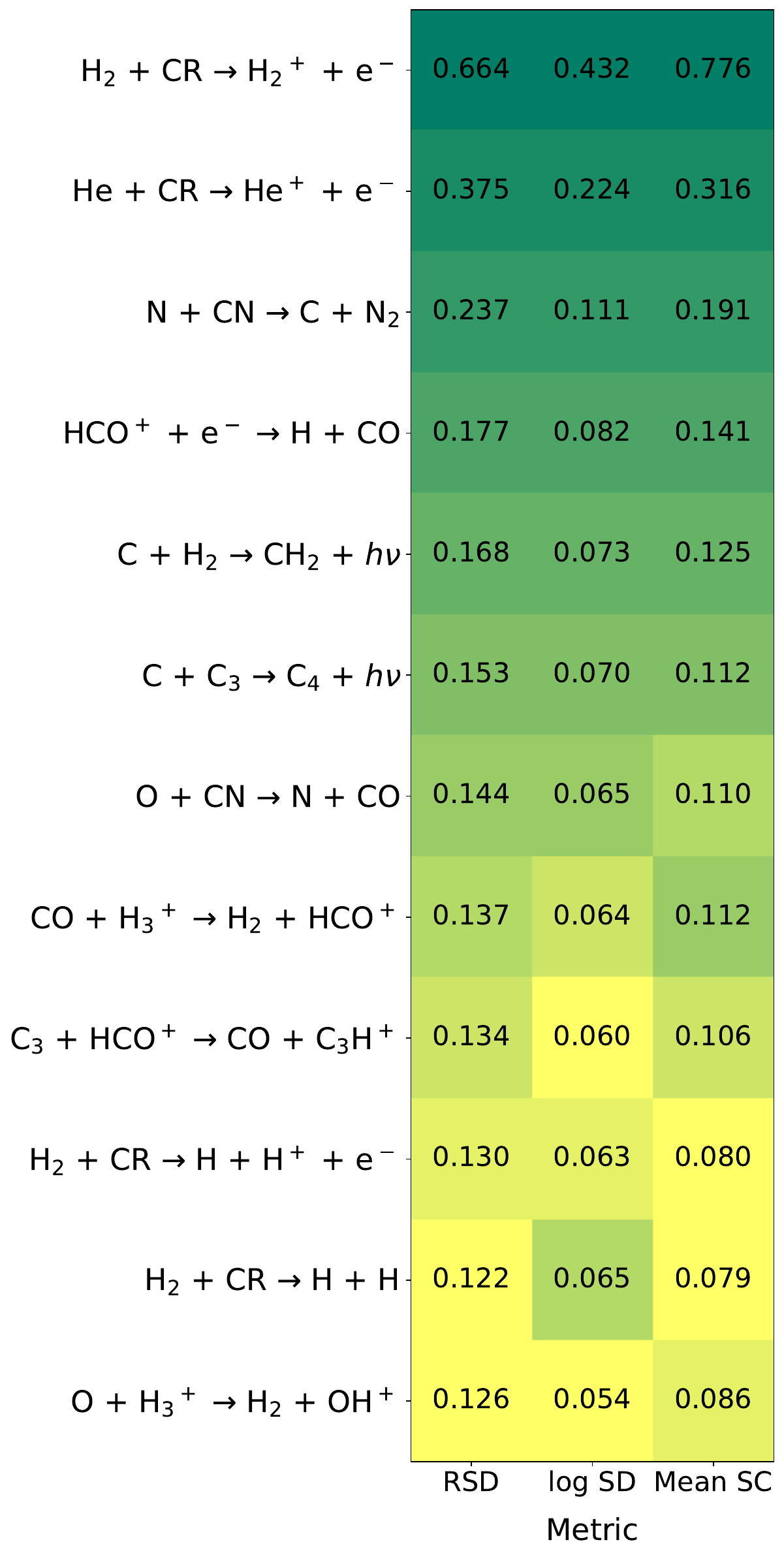}\hfill
    \includegraphics[width=.61\textwidth]{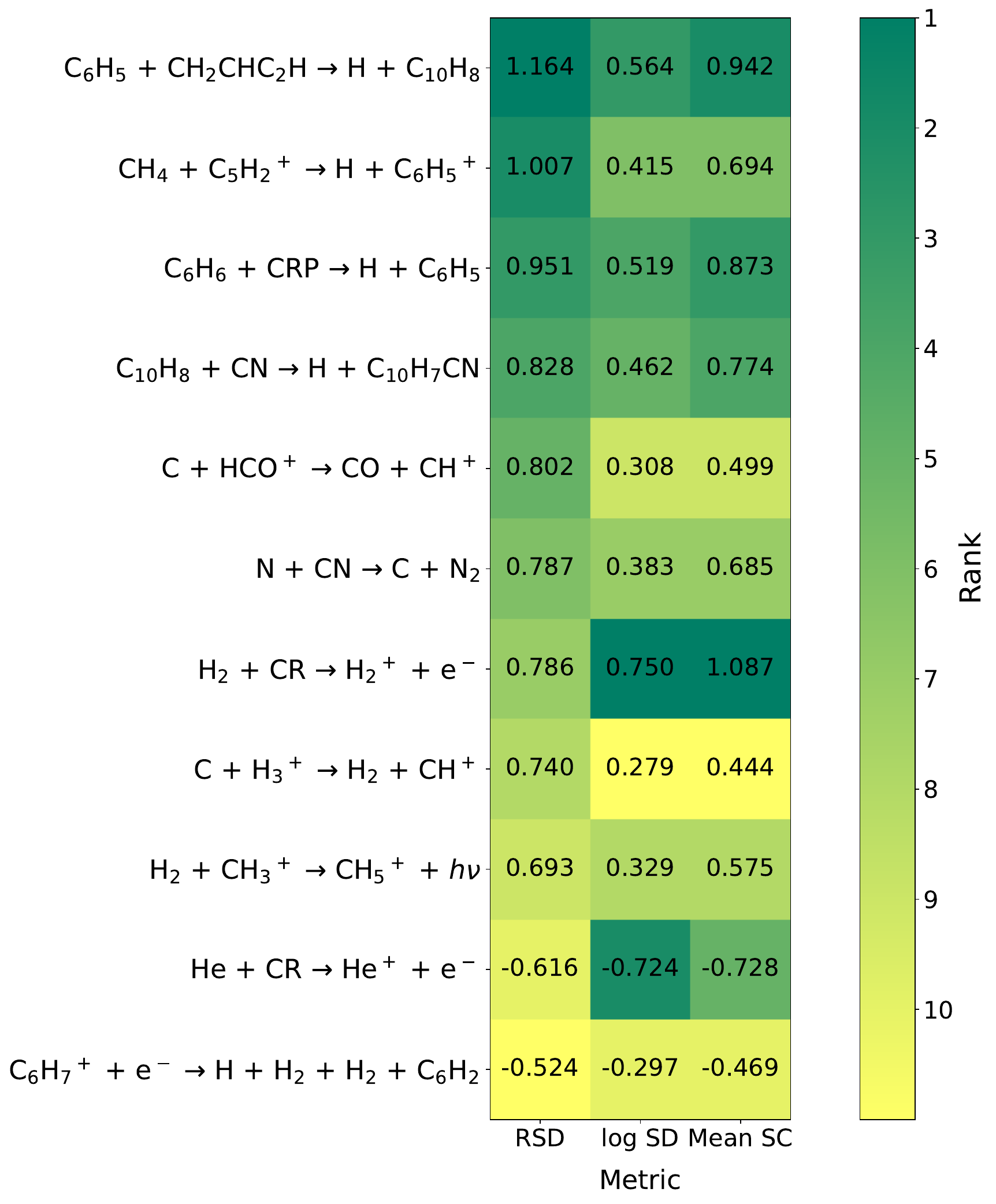}\hfill
    \caption{Heat map of key reactions according to the three OAT metrics at 541100 years for (left) the overall network and (right) \ce{C10H7CN} with a factor of 10 range in variation. All reaction within the top 10 with any of the three metrics are shown. The shading of each cell corresponds to the reaction's rank, with the exact sensitivity metric also given. Reactions ranked \#11 or lower are given the lightest shading.}
    \label{fig:1mag_hmaps}
\end{figure*}

In Figure~\ref{fig:1mag_hmaps} the results of an additional one-at-a-time analysis with a factor of 10 variation are shown. This analysis was performed identically to the factor of 2 variation analysis, but with 20 change factors equally spaced in log scale between 0.1 and 10. These key reaction lists agree well with those obtained with a factor of 2 variation, with slight differences in ordering as well as a few new reactions. In particular, cosmic-ray dissociation and dissociative ionization of \ce{H2} become important to the overall network and the cosmic-ray ionization of He becomes important to \ce{C10H7CN} with a factor of 10 variation in rate coefficient. Additionally, the agreement between the three OAT metrics worsens with this larger range of variation, particularly for the key reactions of \ce{C10H7CN}.

\begin{figure}
    \centering
    \includegraphics[scale=0.7]{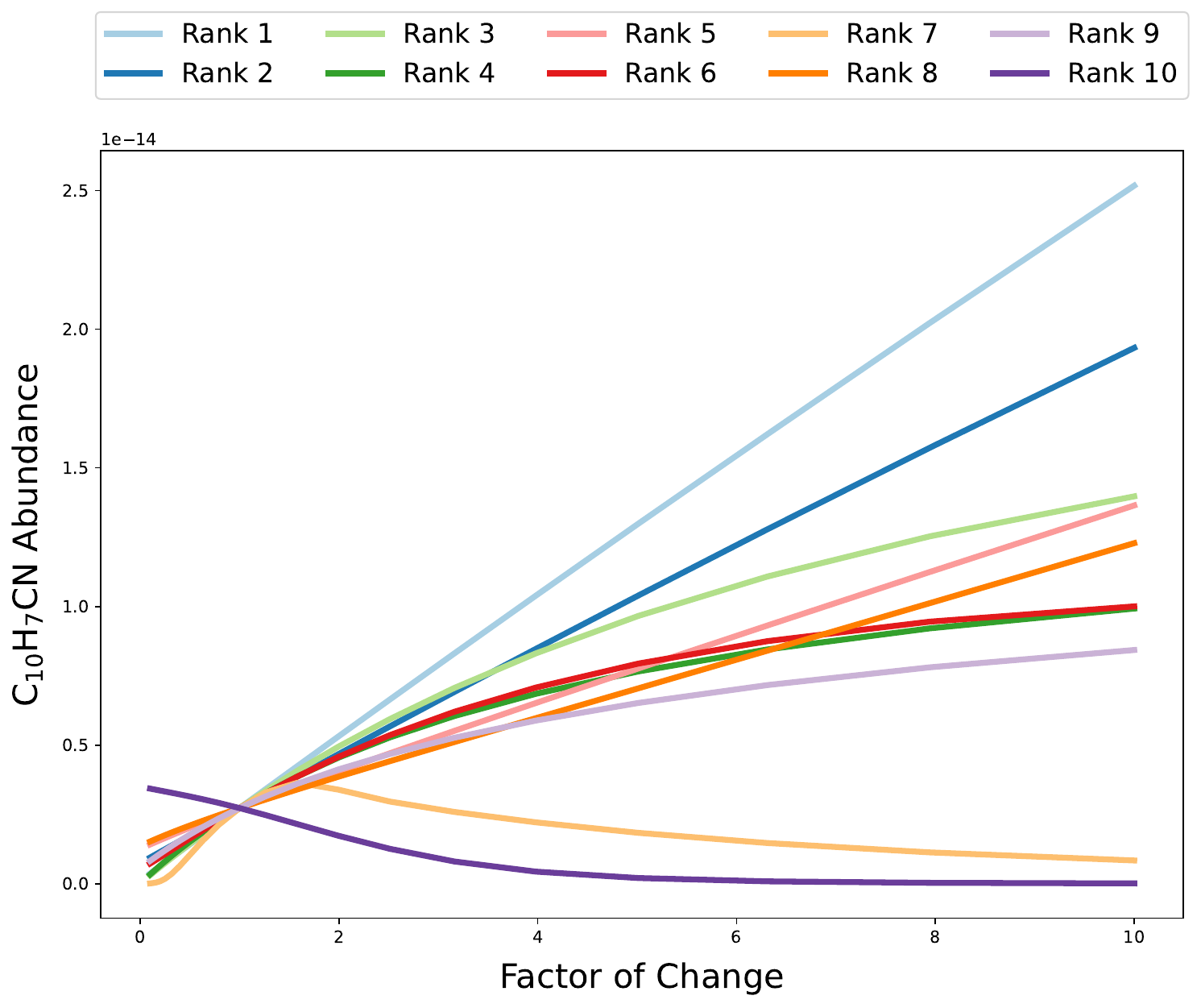}
    \caption{Abundances of \ce{C10H7CN} at 541100 years as a function of factor of change in rate constant for the top 10 reactions. Rankings are based on the RSD metric using the factor of 10 OAT data. Each curve represents one reaction.}
    \label{fig:C10H7CN_OAT_1mag}
\end{figure}

In Figure~\ref{fig:C10H7CN_OAT_1mag}, abundances of \ce{C10H7CN} at 541100 years as a function of change in rate coefficient are shown for its key reactions according to this analysis and the RSD metric. The top 2 reactions, \ce{C6H5 + CH2CHC2H -> H + C10H8} and \ce{CH4 + C5H2+ -> H + C6H5+}, appear to have a roughly linear relationship between \ce{C10H7CN} abundance and factor of change in rate coefficient. Likewise, similar behavior is observed at larger factors of change for the rank 5 and 8 reactions, the reactions of atomic carbon with \ce{HCO+} and \ce{H3+} respectively. The remaining reactions primarily display a power law relationship where the change in \ce{C10H7CN} abundance with respect to change in rate coefficient decreases at larger rate coefficient perturbations. The exceptions are the cosmic-ray ionization of \ce{H2} and He, ranks 7 and 10 respectively, which show strong exponential decays and non-monotonic behavior in the case of \ce{H2} ionization.

\clearpage

\section{Convergence of Monte Carlo Results}

\begin{figure}
     \centering
     \begin{subfigure}[t]{0.49\textwidth}
         \centering
         \includegraphics[width=\textwidth]{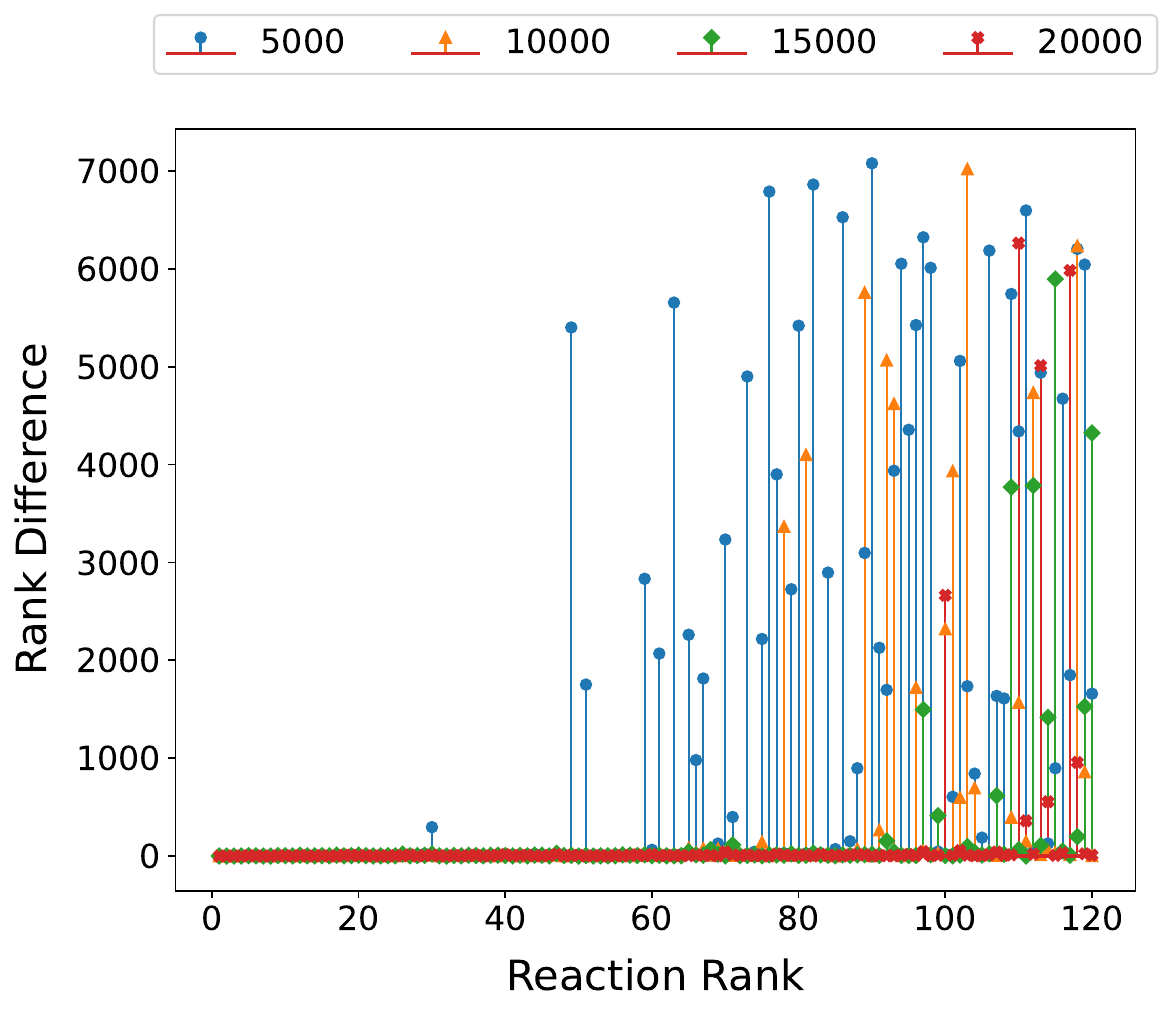}
         \caption{}
     \end{subfigure}
     \hfill
     \begin{subfigure}[t]{0.49\textwidth}
         \centering
         \includegraphics[width=\textwidth]{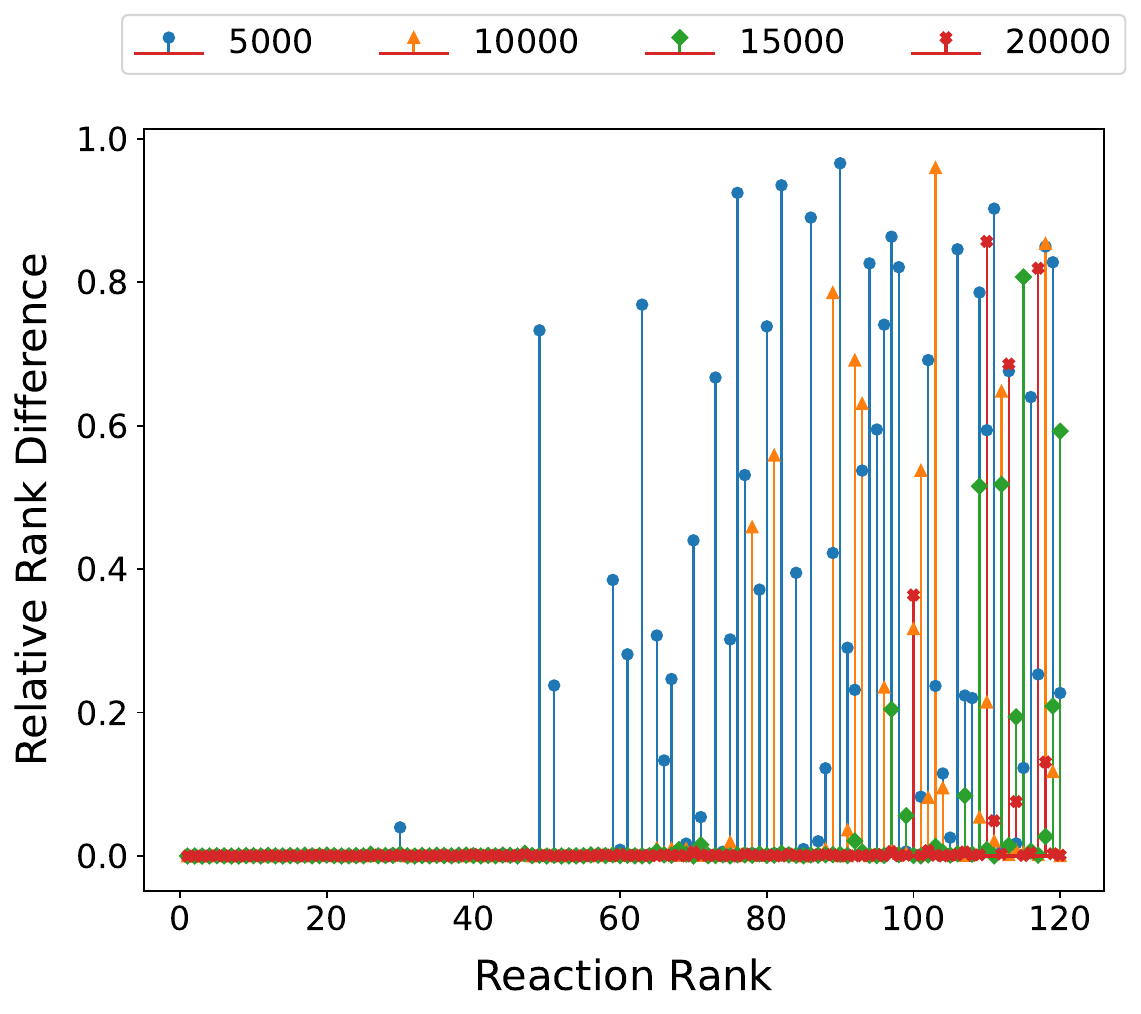}
         \caption{}
     \end{subfigure}
     \hfill
     \begin{subfigure}[b]{0.49\textwidth}
         \centering
         \includegraphics[width=\textwidth]{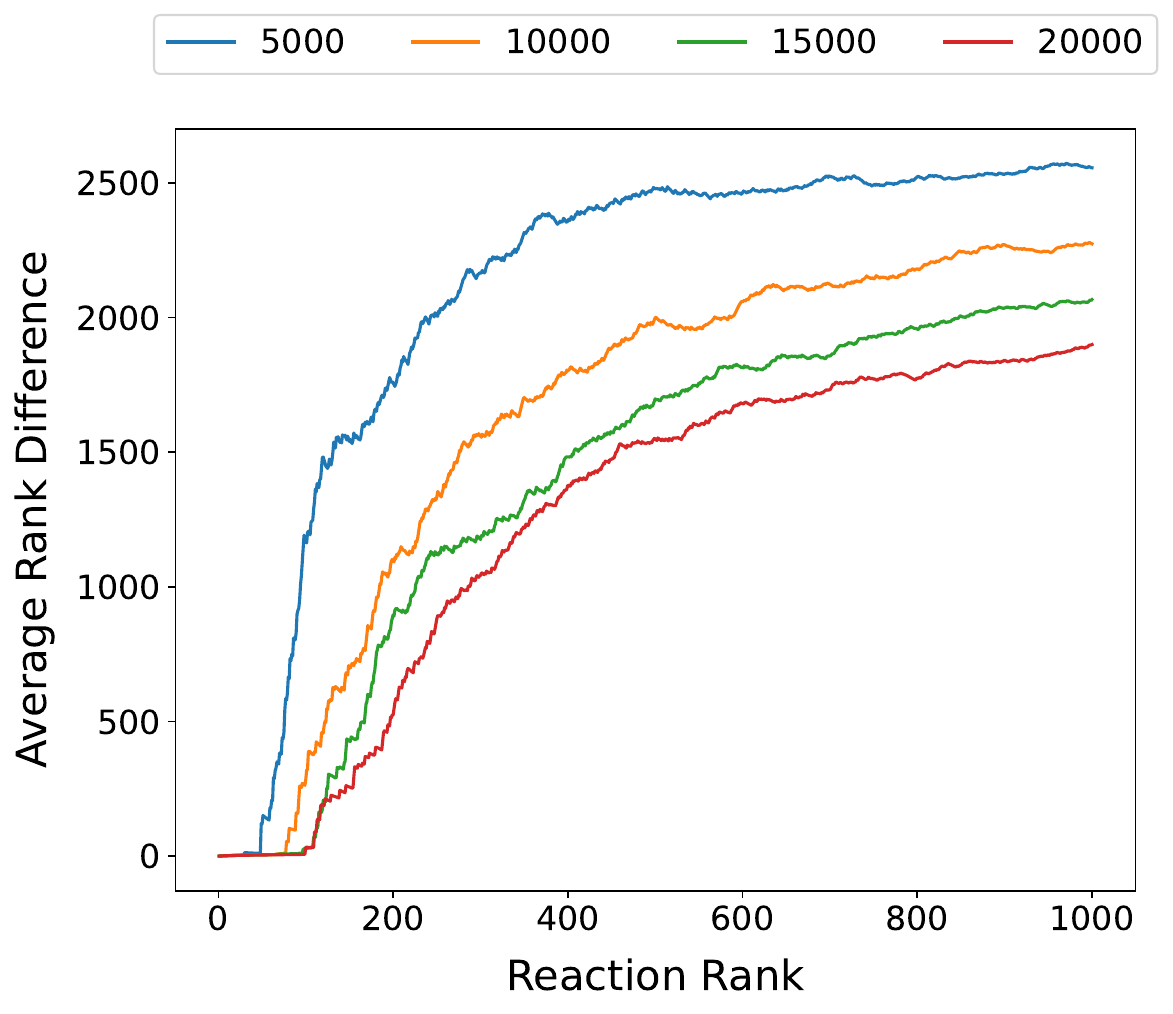}
         \caption{}
     \end{subfigure}
        \caption{Difference in ranking between RSD and RCC metrics as a function of reaction rank, where each color represents a Monte Carlo analysis with some number of iterations. The top two plots are stem plots of the (a) absolute rank difference and (b) relative rank difference, ie. the absolute rank difference divided by the maximum possible rank difference at the given rank. In (c) the rolling average rank difference as a function of rank is plotted.}
        \label{fig:MC_conv}
\end{figure}

In order to investigate the convergence of rank correlations coefficients, four analyses were performed on the same network with 5,000, 10,000, 15,000, and 20,000 iterations. For each analysis all reactions were ranked by average rank correlation coefficients. The differences in these ranks compared to those obtained from the one-at-a-time analysis are displayed in Figure~\ref{fig:MC_conv} as a function of reaction rank. As these methods differ, both in definition of sensitivity as well as number of rate coefficients varied at once, non-zero rank differences are expected. However, this comparison does allow for the determination of when rank correlation coefficients become dominated by noise. This is evidenced by a sudden, large spike in rank difference, followed by consistently large rank differences indicating nearly maximum disagreement. As the number of iterations increases, the point at which this occurs gets pushed back slightly, from rank $\sim$50 for 5,000 iterations to rank $\sim$100 for 20,000 iterations. Likewise, a plot of average rank difference as a function of reaction shows that the steep increase in average rank difference occurs at later ranks when the number of iterations increases. It is clear that even 20,000 iterations is not enough for complete convergence due to the large number of reactions, and thus large parameter space. The key reactions are unaffected, but care should be taken with results beyond these.

\clearpage

\section{Time Dependence of Key Intermediates}
\begin{figure}[t]
    \centering
    \includegraphics[width=\textwidth]{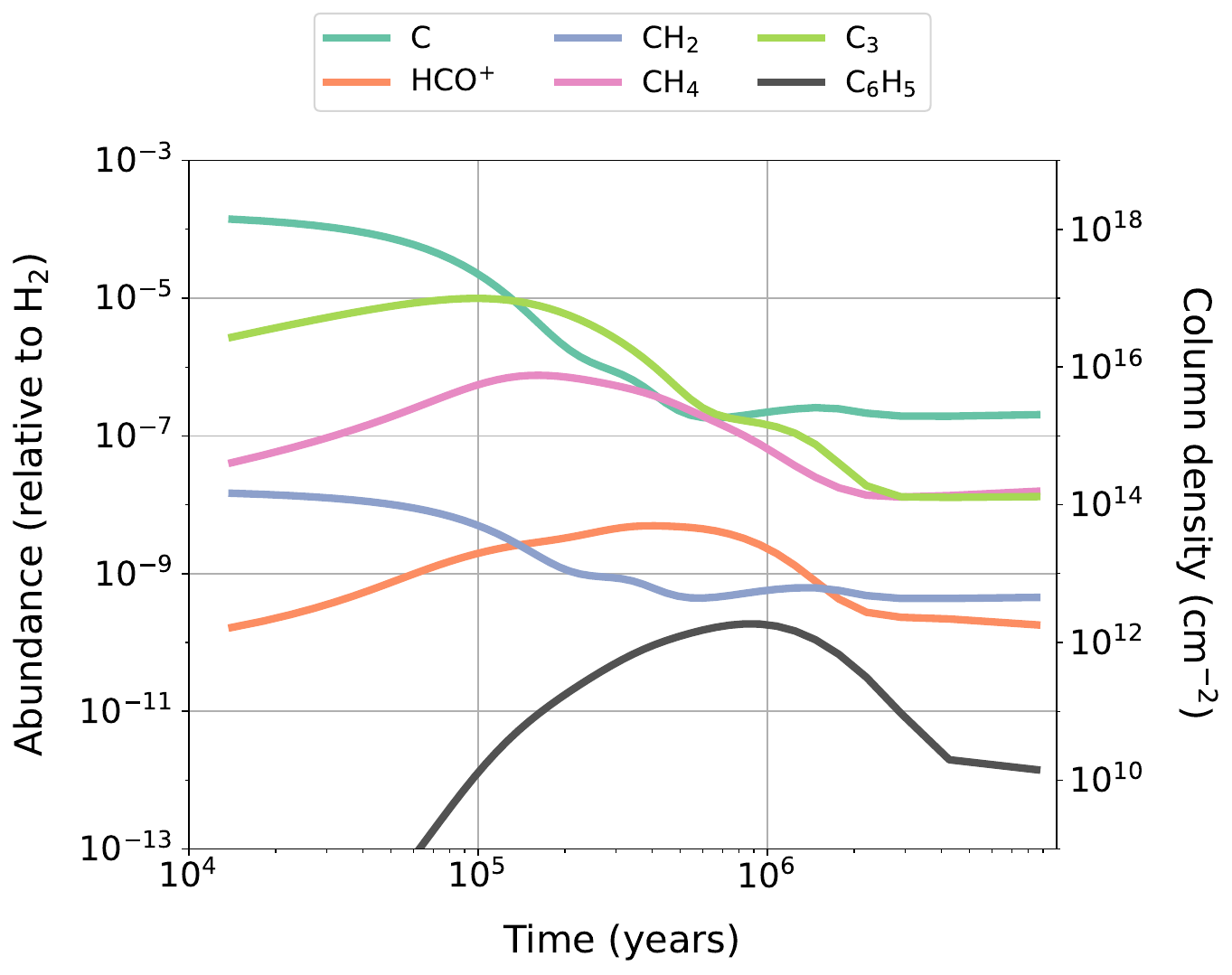}
    \caption{Modeled abundances with respect to \ce{H2} as a function of time for select intermediate species identified as important to the overall network or \ce{C10H7CN} by the sensitivity analyses.}
    \label{fig:intermediates}
\end{figure}

The key reactions for the overall network contain a number of processes that are important for ``early-stage'' hydrocarbon growth, while \ce{C10H7CN} is highly sensitive to many of these same reactions as well as a few directly involving aromatic species. Species such as \ce{C}, \ce{HCO+}, \ce{CH2}, \ce{C3} are important intermediates to hydrocarbon growth that appear in a number of these key reactions. Likewise, species such as \ce{CH4} and \ce{C6H5} play major roles in the formation of large aromatic molecules such as \ce{C10H7CN}. The abundances of these intermediate species in the nominal model are plotted as a function of time in Figure~\ref{fig:intermediates}. C and \ce{CH2} are most abundant before $10^5$ years, after which there is a steep drop in atomic carbon abundance as it is rapidly converted into more complex species. Simpler hydrocarbons and carbon-chain species like \ce{CH4} and \ce{C3} reach their maximum abundances shortly after $10^5$ years but display large abundances throughout the $10^5$-$10^6$ years range. Large, complex hydrocarbons such as \ce{C6H5} do not achieve appreciable abundances until approximately $10^5$ years, with maximum abundances not obtained until \num{5e5} years or later. \ce{HCO+} does not show significant deviations in abundance between $10^5$ and $10^6$ years, consistent with its role as a major proton donor over this range of times.

\clearpage

\section{Rate Coefficients of Key Reactions}
Table~\ref{tab:rate_coeffs} displays the rate coefficient information for the key reactions identified in Figures 3 and 6 of the main text. The \texttt{NAUTILUS} code uses three parameters to calculate rate coefficients at a given temperature. These parameters are given along with the corresponding rate coefficient at 10 K for each reaction. Reactions involving cosmic-ray processes, colored red, are calculated using the equation
\begin{equation}
    k = \alpha\zeta,
    \label{eqa:CR-rates}
\end{equation}
where $\zeta$ is the cosmic-ray ionization rate (set to be \SI{1.3e-17}{\per\second} in our model) and $\alpha$ is a unitless prefactor. Reactions colored blue are ion-neutral processes with rate coefficients calculated using a Su-Chesnavich expression,
\begin{equation}
    k = \alpha\beta\left(1 + 0.0967\gamma\left(\dfrac{300}{T}\right)^{1/2} + \dfrac{300\gamma^2}{10.526T}\right).
    \label{eqa:ionpol2}
\end{equation}
Here $\alpha$ is the branching ratio, $\beta$ is the Langevin rate in units of \SI{}{\cubic\centi\metre\per\second}, and $\gamma$ is a unitless parameter that describes the temperature dependence based on the dipole moment and polarizability of the neutral species. The remaining reactions are calculated using a modified Arrhenius equation,
\begin{equation}
    k = \alpha\left(T/300\right)^{\beta}e^{-\gamma/T},
    \label{eqa:mod-arrhenius}
\end{equation}
where $\alpha$ is the prefactor in units of \SI{}{\cubic\centi\metre\per\second}, $\beta$ is a unitless temperature dependence parameter, and $\gamma$ is the activation energy in units of K.

\begin{table}[htbp]
    \centering
    \caption{\textbf{Key Reaction Rate Parameters}}
    \label{tab:rate_coeffs}
    \centering
    \begin{threeparttable}
    \centering
    \begin{tabularx}{0.75\textwidth}{@{}
    l
    S[table-format=1.2e-2, table-alignment-mode=none]
    S[table-format=-1.2, table-alignment-mode=none]
    S[table-format=4, table-alignment-mode=none]
    S[table-format=1.2e-2]}
    \toprule
    {Reactions} & {$\alpha$} & {$\beta$} & {$\gamma$} & {k (10 \unit{\kelvin})}\\
    \midrule
    \color{red}{{\ce{H2 + CR -> H2+ + e-}}} & \textcolor{red}{\num{9.30e-01}} & \color{red}{0} & \color{red}{0} & \color{red}{\num{1.21e-17}} \\
    {\ce{C + H2 -> CH2 + {h\nu}}} & 1.00e-17 & 0 & 0 & 1.00e-17 \\
    {\ce{C + C3 -> C4 + {h\nu}}} & 4.00e-14 & -1.00 & 0 & 1.20e-12 \\
    \color{red}{{\ce{C + CRP -> C+ + e-}}} & \color{red}{\num{1.02e+03}} & \color{red}{0} & \color{red}{0} & \color{red}{\num{1.33e-14}} \\
    \color{red}{{\ce{He + CR -> He+ + e-}}} & \color{red}{\num{5.00e-01}} & \color{red}{0} & \color{red}{0} & \color{red}{\num{6.50e-18}} \\
    {\ce{C + CH2 -> H + CCH}} & 1.00e-10 & 0 & 0 & 1.00e-10 \\
    {\ce{C + H3+ -> H2 + CH+}} & 2.00e-09 & 0 & 0 & 2.00e-09 \\
    {\ce{C + HCO+ -> CO + CH+}} & 1.10e-09 & 0 & 0 & 1.10e-09 \\    
    {\ce{HCO+ + e- -> H + CO}} & 2.80e-07 & -0.69 & 0 & 2.93e-06 \\
    {\ce{C + C5 -> C3 + C3}} & 1.50e-10 & 0 & 0 & 1.50e-10 \\
    {\ce{C3 + HCO+ -> CO + C3H+}} & 1.40e-09 & 0 & 0 & 1.40e-09 \\
    {\ce{N + CN -> C + N2}} & 8.80e-11 & 0.42 & 0 & 2.11e-11 \\
    {\ce{O + CN -> N + CO}} & 5.00e-11 & 0 & 0 & 5.00e-11 \\
    \color{blue}{{\ce{CO + H3+ -> H2 + HCO+}}} & \color{blue}{\num{9.45e-1}} & \color{blue}{\num{1.99e-09}} & \color{blue}{0.251} & \color{blue}{\num{2.47e-09}} \\
    {\ce{HCNH+ + e- -> H + HCN}} & 9.62e-08 & -0.65 & 0 & 8.78e-07 \\
    {\ce{O + H3+ -> H2 + OH+}} & 7.98e-10 & -0.16 & 1.41 & 1.18e-09 \\
    {\ce{C4 + e- -> C4- + {h\nu}}} & 1.10e-08 & -0.50 & 0 & 6.02e-08 \\
    \color{red}{{\ce{C6H6 + CRP -> H + C6H5}}} & \color{red}{\num{2.93e+03}} & \color{red}{0} & \color{red}{0} & \color{red}{\num{3.80e-14}} \\
    {\ce{C10H8 + CN -> H + C10H7CN}} & 1.50e-10 & 0 & 0 & 1.50e-10 \\
    {\ce{C6H5 + CH2CHC2H -> H + C10H8}} & 2.50e-10 & 0 & 0 & 2.50e-10 \\    
    {\ce{CH4 + C5H2+ -> H + C6H5+}} & 8.00e-10 & 0 & 0 & 8.00e-10 \\
    {\ce{H2 + CH3+ -> CH5+ + {h\nu}}} & 3.78e-16 & -2.30 & 21.5 & 1.10e-13 \\
    {\ce{H + CH -> C + H2}} & 1.24e-10 & -0.26 & 0 & 5.12e-11 \\
    {\ce{C6H7+ + e- -> H + H2 + H2 + C6H2}} & 2.00e-07 & -0.83 & 0 & 3.37e-06 \\
    \bottomrule
    \end{tabularx}
    \end{threeparttable}
\end{table}

\end{appendices}

\end{document}